\documentclass[journal=jacsat,manuscript=communication,etalmode=truncate,keywords=true]{achemso}
\setkeys{acs}{articletitle=true,maxauthors=10,etalmode=truncate,keywords=true}
\usepackage{amsmath,amssymb,units,txfonts}
\usepackage{amsfonts,bm}
\usepackage{graphicx,helvet}
\usepackage[bookmarks=true,colorlinks=true,urlcolor=blue,linkcolor=blue,citecolor=blue]{hyperref} 
\usepackage[usenames,dvipsnames]{color}
\usepackage{multirow}
\usepackage{colortbl,cuted}

\definecolor{JPCCBlue}{RGB}{34,80,169}
\definecolor{abstractcolor}{RGB}{255,243,201}
\definecolor{abstractcolor}{RGB}{255,243,201}
\makeatletter\newenvironment{abstractbox}{%
   \begin{lrbox}{\@tempboxa}\begin{minipage}{0.988\textwidth}}{\end{minipage}\end{lrbox}%
   \colorbox{abstractcolor}{\usebox{\@tempboxa}}
}\makeatother

\renewcommand*\authorfont{\flushleft}
\renewcommand*\affilfont{\mdseries\flushleft\textrm}
\renewcommand*\titlesize{\Large}
\renewcommand*\titlefont{\flushleft\sf\bfseries}
\def\refname{\sffamily\color{JPCCBlue}\Large$\blacksquare$\normalsize\ REFERENCES}
\usepackage{titlesec}
\titleformat{\section}{\bfseries\sffamily\color{JPCCBlue}}{\thesection.~}{0pt}{}
\titleformat{\subsection}[runin]{\bfseries\sffamily\normalsize}{\indent\thesubsection.~}{0pt}{}[.]
\titlespacing{\subsection}{0pt}{0pt}{*1}
\titleformat{\subsubsection}{\bfseries\sffamily\normalsize}{\thethesubsection.~}{0pt}{}
\titlespacing{\subsubsection}{0pt}{0pt}{*0}

\title{Understanding Charge Transfer in Donor--Acceptor/Metal Systems: A Combined Theoretical and Experimental Study}
\author{J.~L.~Cabellos}
\affiliation[Nano-Bio Spectroscopy Group]{\footnotemark[2]{\ } Nano-Bio Spectroscopy Group and ETSF Scientific Development Center, Departamento de F{\'{\i}}sica de Materiales, Universidad del Pa{\'{\i}}s Vasco UPV/EHU, E-20018 San Sebasti\'{a}n, Spain}
\alsoaffiliation[Donostia International Physics Center]{\newline\footnotemark[3]{\ } Donostia International Physics Center (DIPC), Paseo de Manuel de Lardizabal 4, E-20018 San Sebasti\'{a}n, Spain}
\author{D.~J.~Mowbray}
\email{Duncan.Mowbray@gmail.com}
\affiliation[Nano-Bio Spectroscopy Group]{\footnotemark[2]{\ } Nano-Bio Spectroscopy Group and ETSF Scientific Development Center, Departamento de F{\'{\i}}sica de Materiales, Universidad del Pa{\'{\i}}s Vasco UPV/EHU, E-20018 San Sebasti\'{a}n, Spain}
\alsoaffiliation[Donostia International Physics Center]{\newline\footnotemark[3]{\ } Donostia International Physics Center (DIPC), Paseo de Manuel de Lardizabal 4, E-20018 San Sebasti\'{a}n, Spain}
\author{E.~Goiri}
\affiliation[Donostia International Physics Center]{\newline\footnotemark[3]{\ } Donostia International Physics Center (DIPC), Paseo de Manuel de Lardizabal 4, E-20018 San Sebasti\'{a}n, Spain}
\author{A.~El-Sayed}
\affiliation[Aplicada I]{\newline\footnotemark[5]{\ } Departamento de F{\'{\i}}sica Aplicada I, Universidad del Pa{\'{\i}}s Vasco UPV/EHU, E-20018 San Sebasti\'{a}n, Spain}
\author{L.~Floreano}
\affiliation[Laboratorio Nazionale TASC]{\newline\footnotemark[4]{\ } CNR-IOM, Laboratorio TASC, Basovizza SS-14 Km 163.5, I-34149 Trieste, Italy}
\author{D.~G.~de Oteyza}
\affiliation[U.C.~Berkeley]{\newline\footnotemark[6]{\ } Department of Physics, University of California at Berkeley, Berkeley, California 94720, USA}
\alsoaffiliation[Centro de F{\'{\i}}sica de Materiales]{\newline$^\perp$Centro de F{\'{\i}}sica de Materiales CSIC/UPV-EHU-Materials Physics Center, Paseo de Manuel de Lardizabal 5, E-20018 San Sebasti\'{a}n, Spain}
\author{C.~Rogero}
\affiliation[Donostia International Physics Center]{\newline\footnotemark[3]{\ } Donostia International Physics Center (DIPC), Paseo de Manuel de Lardizabal 4, E-20018 San Sebasti\'{a}n, Spain}
\alsoaffiliation[Centro de F{\'{\i}}sica de Materiales]{\newline$^\perp$Centro de F{\'{\i}}sica de Materiales CSIC/UPV-EHU-Materials Physics Center, Paseo de Manuel de Lardizabal 5, E-20018 San Sebasti\'{a}n, Spain}
\author{J.~E.~Ortega}
\affiliation[Donostia International Physics Center]{\newline\footnotemark[3]{\ } Donostia International Physics Center (DIPC), Paseo de Manuel de Lardizabal 4, E-20018 San Sebasti\'{a}n, Spain}
\affiliation[Aplicada I]{\newline\footnotemark[5]{\ } Departamento de F{\'{\i}}sica Aplicada I, Universidad del Pa{\'{\i}}s Vasco UPV/EHU, E-20018 San Sebasti\'{a}n, Spain}
\alsoaffiliation[Centro de F{\'{\i}}sica de Materiales]{\newline$^\perp$Centro de F{\'{\i}}sica de Materiales CSIC/UPV-EHU-Materials Physics Center, Paseo de Manuel de Lardizabal 5, E-20018 San Sebasti\'{a}n, Spain}
\author{A.~Rubio}
\affiliation[Nano-Bio Spectroscopy Group]{\footnotemark[2]{\ } Nano-Bio Spectroscopy Group and ETSF Scientific Development Center, Departamento de F{\'{\i}}sica de Materiales, Universidad del Pa{\'{\i}}s Vasco UPV/EHU, E-20018 San Sebasti\'{a}n, Spain}
\alsoaffiliation[Donostia International Physics Center]{\newline\footnotemark[3]{\ } Donostia International Physics Center (DIPC), Paseo de Manuel de Lardizabal 4, E-20018 San Sebasti\'{a}n, Spain}
\alsoaffiliation[Centro de F{\'{\i}}sica de Materiales]{\newline$^\perp$Centro de F{\'{\i}}sica de Materiales CSIC/UPV-EHU-Materials Physics Center, Paseo de Manuel de Lardizabal 5, E-20018 San Sebasti\'{a}n, Spain}
\email{Angel.Rubio@ehu.es}

\begin{document}

\begin{strip}
\vspace{-1cm}

\noindent{\color{JPCCBlue}{\rule{\textwidth}{0.5pt}}}
\begin{abstractbox}
\begin{tabular*}{17cm}{b
{7.655cm}r}
\textbf{\color{JPCCBlue}{ABSTRACT:}}
We develop an effective potential approach for assessing the flow of charge within a two-dimensional donor--acceptor/metal network based on core-level shifts.  To do so, we perform both density functional theory (DFT) calculations and x-ray photoemission spectroscopy (XPS) measurements of the core-level shifts for three different monolayers adsorbed on a Ag substrate. Specifically, we consider perfluorinated pentacene (PFP), copper phthalocyanine (CuPc) and their 1:1 mixture (PFP+CuPc) adsorbed on Ag(111).
&\includegraphics[width=9cm]{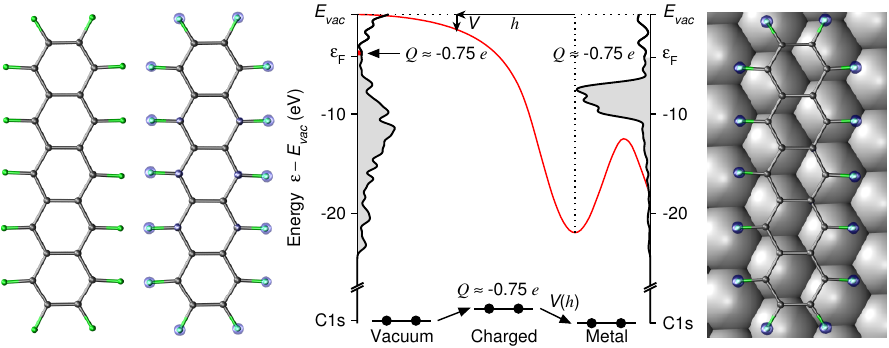}\\
\multicolumn{2}{p{17cm}}{
\noindent{{\color{JPCCBlue}{\textbf{KEYWORDS:}}} PFP, CuPc, Ag(111), XPS, DFT}
}
\end{tabular*}
\end{abstractbox}
\noindent{\color{JPCCBlue}{\rule{\textwidth}{0.5pt}}}
\end{strip}

\section{INTRODUCTION}

X-ray photoemission spectroscopy (XPS) is the most powerful technique to track the changes in the chemical environment of an atom through its core-level shifts. Core-level peaks are readily detected and identified, and core-level shifts can be determined, nowadays, with the highest precision.

In reality, two phenomena contribute to the energy shift of a core-level during a chemical process. First, there is the change in the number of valence electrons or charge transfer into the atom or molecule, which is the quantity we wish to determine. Second, and sometimes overlooked, is the change in the way all energy levels in each atomic species, including core-electron levels, are screened by the external environment. This may include screening from the surrounding atoms, molecules or the substrate. As these two contributions are often of the same order of magnitude, and typically differ in sign, even a substantial charge transfer may result in quite a small core-level shift \cite{Ortega09FrozenLayer,Ortega09Assemblies}.

Experimentally, core-level shifts are known to be affected by the so-called photoemission final-state effects, i.e., the changes in the screening of the photoemission core-hole. Moreover, core-hole screening can vary so much, e.g., during the oxidation of a metal \cite{Ortega94}, as to basically dictate the core-level shift. At this point, only photoemission theory can be used to account for both the (initial state) core-electron and the (final state) core-hole screening, in order to assess the charge transfer from the (experimental) XPS core-level shift. However, this requires expensive calculations of the photoemission excited state 
\cite{1cls,clsInteraction,Cole97,triguero98,Faulkner98CLS,Cole02CLSXPS,Stierle03CLSNiAl,Alagia05,Olovsson06CLSRev,Olovsson10CLS,Li10CLSmolads,Schmidt10} even for the simplest system, and hence it becomes unaffordable for molecular complexes with large numbers of atoms.

\renewcommand{\thefootnote}{}
\footnotetext{\hspace{-0.5cm}This document is the unedited Author's version of a Submitted Work that was subsequently accepted for publication in The Journal of Physical Chemistry C, copyright \copyright American Chemical Society after peer review. To access the final edited and published work see \href{http://dx.doi.org/10.1021/jp3004213}{http://dx.doi.org/10.1021/jp3004213}.}

Systems where the changes in core-hole screening are small are better suited for combined theory/experiment studies, since final state effects can be neglected. For example, when moving from a pure to a mixed donor/acceptor molecular monolayer on the same metal substrate. These systems have stimulated much interest for both experimentalists and theorists, particularly for olefins
\cite{Dai05OlefinsAg111},
 pentacene
\cite{Louie02PENGrowth,Endres04PEN,HighPentaceneLDA,Cantrell08PENDiff,Koch08PENAg111,Toyoda10PENCuAgAu111,Mete10PENAg111,Oteyza10Fluorination},
perfluorinated pentacene
\cite{Sakurai09PFPBi,Duhm2010PFPAg111,Witte11PFPAg111},
their 1:1 mixture
\cite{Tokito04PFPPEN,Hinderhofer07PENandPFP,Koch08PENPFP,Toyoda11PENPFPCu111},
copper phthalocyanine
\cite{Gerlach05CuPcAg,Wee09CuPcAg111,Wee09FCuPcAgAu111,Kumpf10CuPcAg111,Moller96CuPcAg111,TMPcAg100},
and fluorinated phthalocyanine
\cite{Wee10FCuPcPEN,PEN+FCuPcAu111,FCuPcPENAu111},
on the (111) facet of the coinage metals. It has been repeatedly reported that such two-dimensional blending of donors and acceptors gives rise to core-level shifts in all atomic components. Here we show that the corresponding transfer of charge can be estimated from the core-level shift if changes in the external environment during the molecular blending process are properly accounted for. In fact, we demonstrate that, in the absence of major chemical disruptions, an effective potential approach can be utilized for a semi-quantitative evaluation of changes in core-electron screening. This effective potential approach is computationally cheap, thereby allowing a fast and accurate determination of molecular charge transfer.

The present work combines density functional theory (DFT) calculations with the XPS study of perfluorinated pentacene (PFP), copper phthalocyanine (CuPc) and their 1:1 mixture (PFP+CuPc), on the (111) surface facet of Ag.  In Section~\ref{sec:methodology} we provide details of the computational and experimental methods employed, a derivation of the effective potential model, and a discussion of the initial state method used to calculate core-level shifts. We also test the reliability of these DFT calculations by a direct comparison of scanning tunneling microscopy (STM) measurements and simulations.  The results are discussed in Section~\ref{sec:resultsanddiscussion}, beginning with the charging of PFP on Ag(111). The initial state method for calculating core-level shifts is then compared with XPS measurements for multilayers and monolayers of PFP on Ag(111).
  The dependence of the calculated core-level shifts on both charge transfer and the external potential is then demonstrated for monolayers of pure PFP, CuPc and their 1:1 mixture PFP+CuPc, and compared with experiment.  These results are then used to compare the calculated charge transfer from DFT with that obtained from a model based on the core-level shifts and change in external potential.  The model is then applied to the XPS core-level shifts, to estimate the experimental charge of PFP on Ag(111). This is followed by concluding remarks in Section~\ref{sec:conclusions}.  

In Appendix~\plainref{sec:stm} we provide provide further details of the STM simulation method employed. We then compare results from three different final state methods, described in Appendix~\plainref{sec:finalstatemethods}, with XPS measurements and initial state method calculations for monolayers of PFP on Ag(111) in Appendix~\plainref{sec:comparison}.

\section{METHODOLOGY}\label{sec:methodology}

\subsection{Computational Methods} DFT calculations have been performed using the real-space projector augmented wavefunction (PAW) \cite{blo} code \textsc{gpaw} \cite{gpaw1,gpaw2}, within the local density approximation (LDA) for the exchange-correlation (xc)-functional \cite{LDA}, using a grid spacing of 0.2~\AA. An electronic temperature of $k_B T \approx 0.1$~eV was employed to obtain the occupation of the Kohn-Sham orbitals, with all energies extrapolated to $T = 0$~K. 

\begin{figure}[!t]
\centering
\includegraphics[width=\columnwidth]{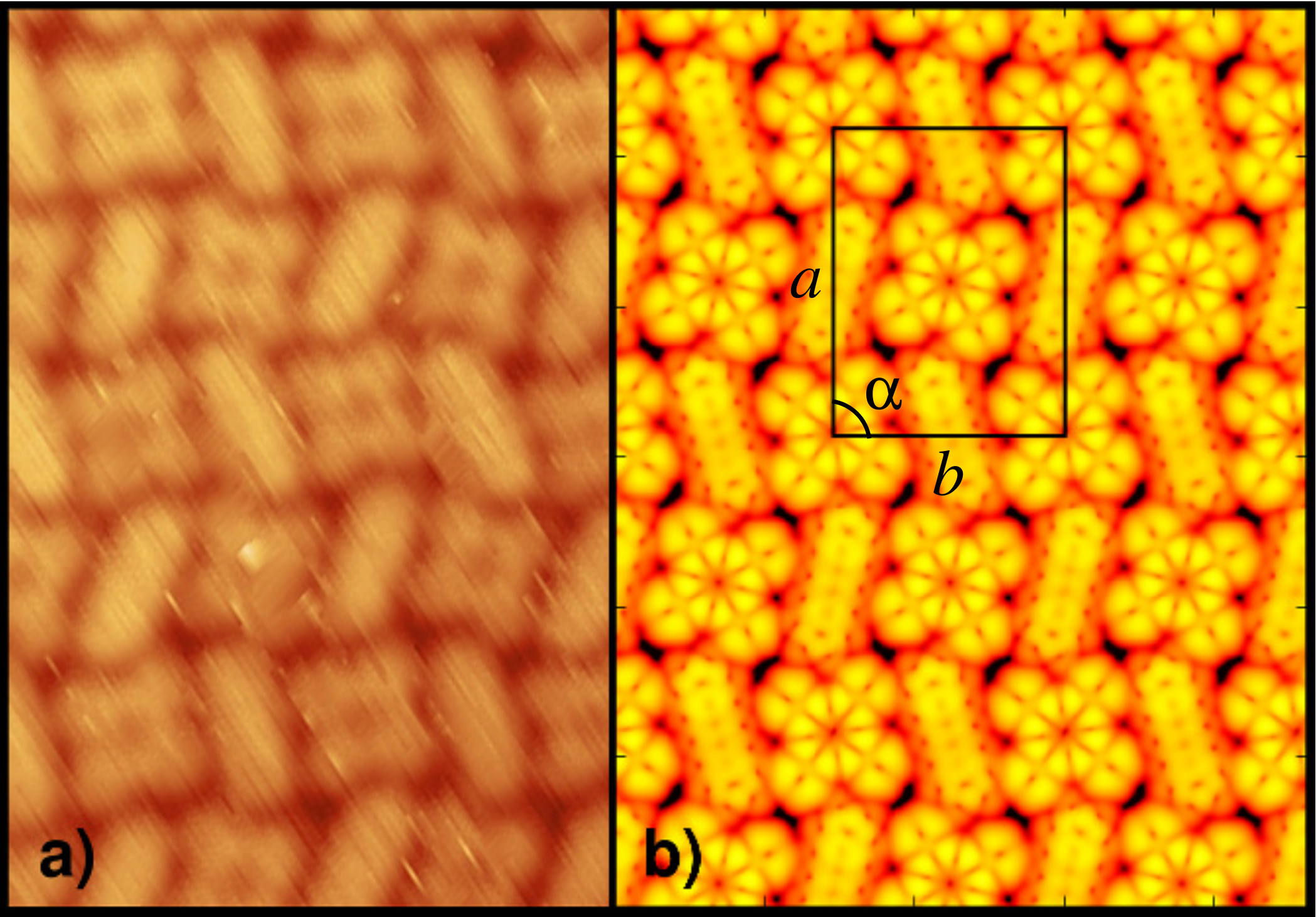}
\caption{STM images of a mixed 1:1 monolayer of CuPc and PFP adsorbed on Ag(111) from (a) experiment \cite{goiri} and (b) a Tersoff-Hamann \cite{stm} calculation. The unit cell of the calculation is also shown, with $a$, $b$, and $\alpha$ provided in Table~\ref{para}.}
\label{fgr:STM}
{\color{JPCCBlue}{\rule{\columnwidth}{1pt}}}
\end{figure}

Monolayers of PFP, CuPc, and their 1:1 mixture PFP+CuPc have been structurally optimized until a maximum force below 0.05~eV/\AA~was obtained in vacuum and adsorbed on the Ag(111) surface, while keeping the coordinates of the metal slab fixed. The lattice parameters, shown in Table~\ref{para}, are those commensurate with the experimental bulk lattice constant for Ag of 4.09~\AA \cite{Ashcroft}, which are nearest to the periodicity of the monolayer on the surface as observed by STM  \cite{goiri}. In Figure~\ref{fgr:STM} (a) and (b) we compare the measured and calculated STM images, respectively, for the mixed 1:1 PFP+CuPc monolayer on Ag(111). The images agree quite well justifying our approach. Further details concerning the STM simulation procedure are provided in Appendix~\plainref{sec:stm}.

 \begin{table}[!t]
\caption{\textrm{
Lattice parameters \textit{a}, \textit{b}, and {$\bm{\alpha}$}, for the PFP and CuPc pure and 1:1 mixed monolayers on Ag(111) as obtained from STM (upper values)\cite{goiri}, and those commensurate with the experimental bulk lattice constant for Ag of 4.09~\AA \cite{Ashcroft}, used in the calculations (lower values)}}\label{para}
\begin{tabular}{lr@{.}lr@{.}ll}
\multicolumn{6}{>{\columncolor[gray]{0.9}}c}{ }\\[-3mm]
\rowcolor[gray]{0.9}monolayer& \multicolumn{2}{>{\columncolor[gray]{.9}}c}{$a$ (\AA)} & \multicolumn{2}{>{\columncolor[gray]{.9}}c}{$b$ (\AA)} & \multicolumn{1}{>{\columncolor[gray]{.9}}c}{$\alpha$}\\[1mm]
\multirow{2}{*}{PFP}      & 17&$0\pm 1.0$ &  8&$8\pm 0.9 $ & $62^\circ\pm 2^\circ$ \\
      & 17&352        &  8&676         & ${60}^\circ$ \\[1mm]
\multirow{2}{*}{CuPc}     & 14&$1\pm 0.8$ & 13&$9\pm 0.7 $ & $88^\circ\pm 4^\circ$ \\
     & 14&460        & 15&028         & ${90}^\circ$ \\[1mm]
\multirow{2}{*}{PFP+CuPc} & 29&$3\pm 0.6$ & 22&$0\pm 2.0 $ & $89^\circ\pm 6^\circ$ \\
 & 30&055        & 23&137         & ${90}^\circ$ \\
\end{tabular}
{\color{JPCCBlue}{\rule{\columnwidth}{1pt}}}
\end{table}

The Ag(111) surface has been modeled using $N=1,2,3,\ldots,6$ layers with the slabs separated from their periodic images by more than 12~\AA\ of vacuum.  For such large unit cells as those used herein, \emph{cf}.~Table~\ref{para}, we found a $\Gamma$ point calculation was sufficient to converge the electronic density for the mixed PFP+CuPc monolayer, while for pure PFP and CuPc we employed a ($1\times3\times1$)  and ($3\times3\times1$) $\textbf{k}$-point sampling, respectively. However, for calculating core-level shifts from an initial or final state method, a $\Gamma$-point calculation based on the optimized geometry was found to be sufficient to converge the core 1s levels.

For each monolayer, both spin polarized and spin paired calculations were performed in vacuum.  For CuPc we find the molecule has a magnetic moment of $\mu = 1\mu_0$ in vacuum.  However, in the 1:1 mixture consisting of two CuPc and two PFP  molecules, shown in Figure~\ref{fgr:STM},
 CuPc is paramagnetic with no net magnetic moment, as was also the case for PFP.  For this reason spin paired calculations were sufficient to describe the adsorption of the two monolayers on the Ag(111) surface.

To model the effective potential for the semi-infinite Ag(111) surface $V$ in the experiment, we have used a fully-relaxed 13 layer Ag(111) slab.  Such a thick slab is required to completely converge the band structure of the Ag(111) surface, and remove surface---surface interactions from the calculation. In this case, we have employed the generalized gradient approximation (GGA) as implemented by Perdew, Burke, and Ernzerhof (PBE) for the xc-functional\cite{PBE} with an optimized bulk lattice constant of 4.166~\AA\ in the surface plane, and a ($11\times11\times1$) $\textbf{k}$-point sampling.  The GGA-PBE xc-functional is expected to provide a more accurate description of the experimentally-observed effective potential, by removing the spurious long range over-binding found in LDA calculations.

\subsection{Experimental Setup}

The Ag crystal was cleaned by cycles of Ar$^{\mathrm{+}}$ ion sputtering followed by annealing to about 400$^\circ$C. Molecule coverage was calibrated using a quartz crystal microbalance. Measurements took place in UHV conditions, with base pressures in the 10$^{-10}$ mbar range.

STM measurements were performed at a commercial Omicron VT-STM in constant current mode with electrochemically etched W tips. The XPS experiments were performed at ALOISA beamline of the Elettra Synchrotron in Trieste, Italy.  A photon energy of 500~eV was used for the C1s and N1s core-levels, and 810~eV was used for the F1s core-level. Cleanliness of the surface was checked by measuring the C1s and O1s spectrum.  At the same time, coverage in pure and mixed layers was verified through analysis of the N1s and F1s core-level intensities, with the Ag3d level as common reference.

\subsection{Theoretical Model}\label{sec:theoreticalmodel}

\begin{figure*}[!t]
\centering
\includegraphics[width=0.9\textwidth]{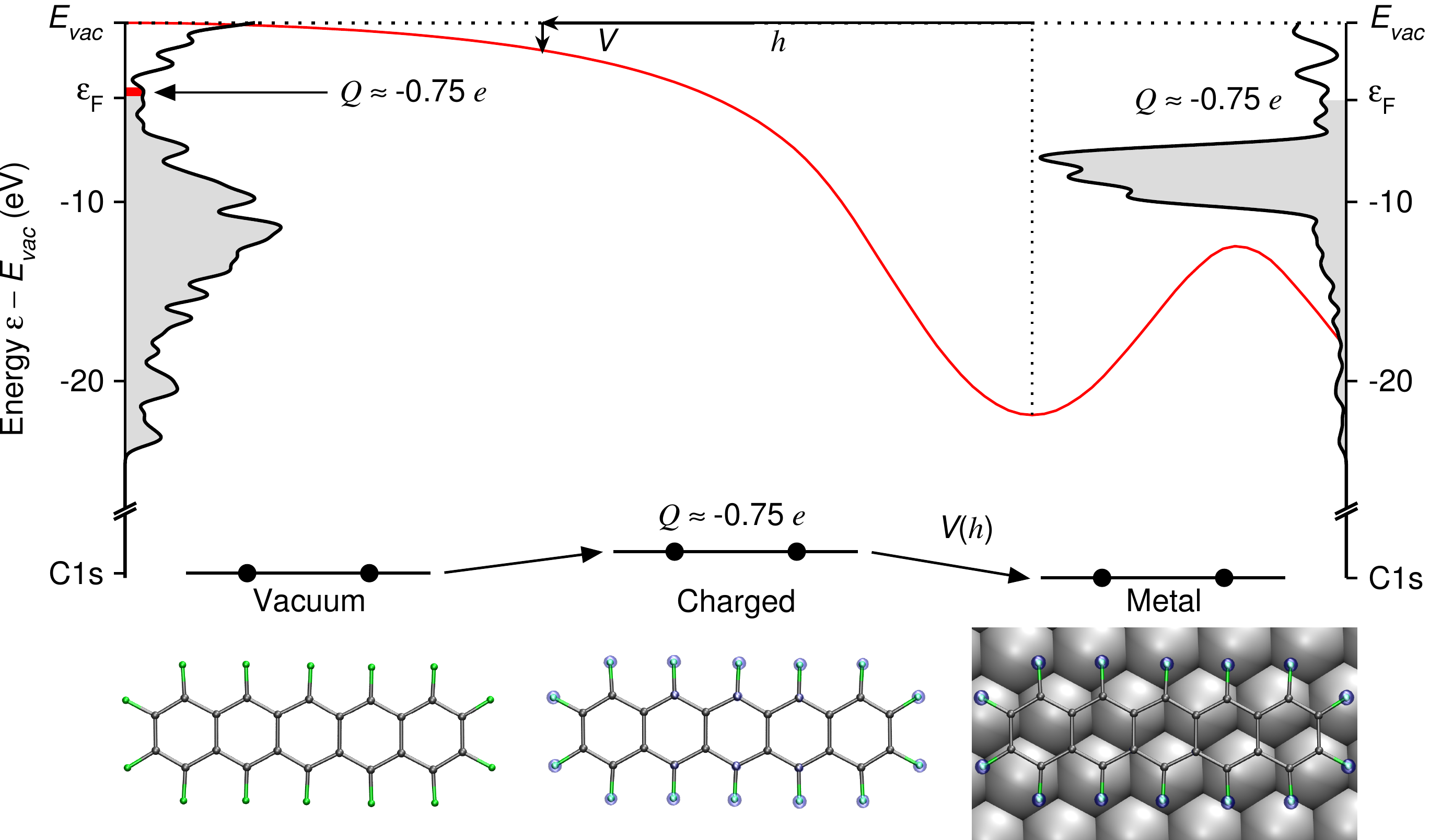}
\caption{Schematic depicting the initial state method for calculating core-level energy shifts as applied to a PFP monolayer in vacuum, charged, and adsorbed on a Ag(111) three layer surface. The calculated density of states (DOS) for the initial state ({\Large{\textbf{------}}}), charge of the molecule $Q$, external potential $V$ ({\color{red}{{---------}}}), and C1s energy shifts are shown.  Occupation of the DOS is denoted by grey regions, with charge added to the molecule in vacuum marked in red.  Depictions of the neutral, charged ($Q \approx -0.75$~$e$), and adsorbed PFP on Ag(111) are also shown below, with the change in charge density depicted by isosurfaces of $-0.1$~$e/a_0^3$.  C and F atoms are depicted by gray and green balls, respectively.  Note  the DOS for the PFP monolayer in vacuum was increased by a factor of five for clarity.}
\label{fgr:initialstatemethod}
{\color{JPCCBlue}{\rule{\textwidth}{1pt}}}
\end{figure*}

Once the equilibrium structures were obtained from standard DFT, calculations of the core-level shifts were then performed.  The core-level shift, \(\Delta E\), is defined as the difference in binding energy of an electron in a core state between two environments 
\(E_{1}\) and \(E_{2}\), so that
\begin{equation}
\Delta E \equiv E_{2} - E_1
\end{equation}
As a particular example, we may consider \(\Delta E\) the difference in binding energy of the C1s level of PFP between a monolayer in vacuum and a mixed 1:1 monolayer of PFP+CuPc adsorbed on the Ag(111) surface.

The core-level shift is determined, to a large extent, by two factors: (1) the charge transfer into the atom $Q_a$, and (2) the effect of screening of the atom by the external environment.  In the linear regime, where major chemical interactions are absent, the change in screening should be related to the change in the effective potential coming from the atom/molecule's external environment \(V\), at the atom's position \(\mathbf{r}_a\).  Under these conditions, the core-level shift should be an additively separable function of the charge transfer and change in external potential, i.e.
\begin{equation}
\Delta E(Q_a, V) \approx f_{\textit{mol}}(Q_a) + g(V)\label{eqn:ECLS}
\end{equation}
Here \(g(V)\) describes the change in screening of the atom between the two 
 environments.  If we assume the molecule does not undergo significant alteration between the two environments, then \(g\) should only weakly depend on the molecule.  In this weak interaction limit, we may further approximate the screening from the environment simply by
\begin{equation}
g(V) \approx - V(\mathbf{r}_a)\label{eqn:gscr}
\end{equation}
In this way, the core-levels of an atom should shift to \emph{stronger} binding energies when the molecule enters a binding external potential, e.g.\ due to a screening by a surface.  Although somewhat drastic, we shall show that this crude approximation captures the physics of the core-electron screening for such systems.

The dependence of the core-level shift on the charge transfer \(f_{\textit{mol}}(Q_a)\) should depend only on the local chemical environment of the molecule, and may be assumed independent of the external environment.  In this way, for typical donor--acceptor charge transfers, \(f_{\textit{mol}}(Q_a) \approx \kappa_a Q_a \approx \kappa_X Q_X\), where $X$ is one of the symmetrically inequivalent chemical environments on the molecule, i.e.\ atomic species, and \(\kappa_{a/X}\) are constants. Further simplification is possible by reformulating \(f_{\textit{mol}}\) in terms of the total charge transfer into the molecule \(Q\).  This is done by assuming that the fraction of the total charge which is given to atomic species $X$, \(\frac{Q_X}{Q}\), is a linear function of the total charge.  More precisely,
\begin{equation}
f_{\textit{mol}}(Q_X) \approx \kappa_X \frac{Q_X}{Q} Q\approx \left(\xi + \zeta Q\right)Q = \xi Q + \zeta Q^2\label{eqn:fmol}
\end{equation}
where \(\xi > 0\).  This implies that the core-levels should shift to \emph{weaker} binding energies when charge is transferred into the molecule.  Further, if $X$ is less electronegative than the other atomic species in the molecule, i.e.\ C relative to N or F, then \(\zeta > 0\). On the other hand, the opposite would be true for more electronegative atomic species, such as N or F relative to C.

Substituting \ref{eqn:gscr} and \ref{eqn:fmol} into \ref{eqn:ECLS}, we obtain a simple expression for the core-level shift,
\begin{equation}
\Delta E \approx \xi Q + \zeta Q^2 - V\label{eqn:DEModel}
\end{equation}
in terms of the molecule's charge $Q$, the effective potential from the external environment $V$, and two molecule dependent parameters $\xi$ and $\zeta$.  

These two parameters may be obtained by performing core-level shift calculations for a charged pure monolayer, as depicted schematically in Figure~\ref{fgr:initialstatemethod}.  In this case, the screening from the external environment does not play a role $g(V) \approx 0$, so that the core level shifts are only dependent on the local chemical environment of the molecule, i.e.\ $\xi$, $\zeta$, and $Q$.  Since the charge of the molecule is specified within the core-level calculation, by performing a few such calculations for various chargings $Q$, one quickly obtains a good estimate for $\xi$ and $\zeta$.

Once one has obtained $\xi$ and $\zeta$, since $f_{\textit{mol}}$ is assumed to be dependent only on the charge of the molecule, one simply needs to calculate the molecule's charge via a Bader analysis in the external environment.  On the other hand, for an estimate of the external potential $V$, one should separately calculate the effective potential from the external environment, e.g.\ a clean metal substrate or only the surrounding molecules in a mixture.  This is depicted schematically in Figure~\ref{fgr:initialstatemethod} for a pure PFP monolayer on three layer Ag(111).
In this case, $V$ is the effective potential from a clean three layer Ag(111) slab at the height $h$ of the PFP monolayer above the surface.  Altogether, this allows one to model the core-level shifts relative to the neutral molecule using \ref{eqn:DEModel}.

Likewise, the total charge transfer \(Q\) into the molecule may be modeled using the core-level shift relative to the neutral molecule in vacuum, \(E - E_{Q=0}\), by
\begin{equation}
Q \approx -\frac{\xi}{2\zeta} + \sqrt{\frac{\xi^2}{4\zeta^2} + \frac{E - E_{Q=0} + V}{\zeta}}\label{eqn:Qmodel}
\end{equation}
This result allows us to formulate the following effective potential approach for describing charge transfer in donor--acceptor/metal systems based on core-level shifts.
\begin{enumerate}
\item Perform a DFT structural relaxation of the neutral pure monolayer in vacuum.
\item Calculate core-level binding energies for the neutral pure monolayer in vacuum $E_{Q=0}$.
\item Calculate core-level shifts for pure monolayers in vacuum with various chargings $Q$, $\Delta E(Q)$.
\item Obtain $\xi$ and $\zeta$ from a quadratic fit to $\Delta E(Q)$.
\item Measure or calculate heights $h$ or positions $\mathbf{r}_a$ of the monolayer on the metal substrate or in a mixed monolayer.
\item Calculate the external potential $V$ from the metal substrate or mixed monolayer at the heights or positions of the atomic species.
\item Measure or calculate core-level binding energies \(E\) for the atomic species on the metal substrate or in the mixed monolayer.
\item Using $\xi$, $\zeta$, $V$, $E$, and $E_{Q=0}$ in \ref{eqn:Qmodel}, estimate the charge of the molecule $Q$.
\end{enumerate}
Thus, in order to determine the total charge of a molecule $Q$, environment 1 should refer to the neutral molecule, e.g.\ a pure monolayer in vacuum, or a multilayer crystal on a surface.

For this reason, we first consider the core-level shifts between a monolayer of PFP in vacuum and adsorbed on the three layer Ag(111) surface. In this case, though, noticeable changes in core-hole screening are expected. 
To assess how much the core-hole screening may contribute to the core-level shifts, we have used both the initial state method, where core-hole screening is neglected, and three types of final state methods, where the core-hole is directly modeled.  Specifically, we modeled the final state within the full core-hole, half core-hole, and screened core-hole approximations, which are described in Appendix~\plainref{sec:finalstatemethods}.  Although each method has its advantages and disadvantages, overall, we find the initial state method provides the best balance of accuracy with computational costs, for the systems we consider herein.

\subsection{Initial State Method}\label{sec:initialstatemethod}

In the initial state method, depicted schematically in Figure~\ref{fgr:initialstatemethod}, the binding energy is described using the Kohn-Sham eigenenergies for the given core-level relative to the vacuum energy.  This requires an additional DFT calculation using the relaxed geometry, which includes the core 1s levels in the valence for all relevant atoms, i.e.\ C, F, and N.  In this way, an all-electron calculation is performed for the entire molecule, within the PAW method, without requiring a finer grid spacing, e.g.\ \(\sim 0.05\)~\AA.  The binding energy of atomic species $X$'s 1s level \(E\), is then modeled by
\begin{equation}
E \approx - \varepsilon_{X\mathrm{1s}} + E_{\textit{vac}}
\end{equation}
where \(\varepsilon_{X\mathrm{1s}}\) is the energy of a local maxima in the total density of the states due to atomic species $X$'s 1s levels, and \(E_{\textit{vac}}\) is the vacuum energy. This is given by the maximum in the surface averaged effective potential,
\begin{equation}
E_{\textit{vac}} = \max_z \iint_{\mathcal{A}} \frac{dx dy}{\mathcal{A}} V(x,y,z)\\
 \approx \lim_{h\rightarrow\infty} V(h)
\end{equation}
where \(\mathcal{A}\) is the the area of the monolayer in the unit cell, $h$ is the height above the surface,  and $V$ is the effective Kohn-Sham potential.  

To summarize, an initial state calculation of the binding energy $E$ for the 1s core level of atomic species $X$ involves the following procedure:
\begin{enumerate}
\item Perform a DFT structural relaxation of the molecular system.
\item Recalculate using all-electron PAW pseudopotentials for the molecule.
\item Use the surface averaged effective potential to calculate the vacuum level $E_{\textit{vac}}$.
\item Obtain the local maxima in the DOS for atomic species $X$'s core level $\varepsilon_{X\mathrm{1s}}$.
\item Calculate the initial state binding energies for atomic species $X$ using $E = -\varepsilon_{X\mathrm{1s}} - E_{\textit{vac}}$.
\end{enumerate}

Although the Kohn-Sham eigenenergies underestimate the experimental binding energies by \(\sim 10\%\), due to error cancellation, the \emph{shifts} in the binding energies are quite accurately described.  This method also has the advantage of calculating the core-level shifts for all atoms in the molecule simultaneously.  For complex systems such as PFP+CuPc with \(\sim 100\) C atoms, this results in a computational advantage of two orders of magnitude over final state methods, where separate calculations are required for each chemical environment, i.e.\ atomic species.

\section{RESULTS AND DISCUSSION}\label{sec:resultsanddiscussion}

\subsection{Charge of PFP on Ag(111)}

In order to estimate the charge transfer to the molecules we have used  the Bader partitioning scheme~\cite{bader}. This method only requires the DFT all-electron density, with the partitioning of the density determined according to its zero-flux surfaces.

\begin{figure}[!ht]
\includegraphics[width=\columnwidth]{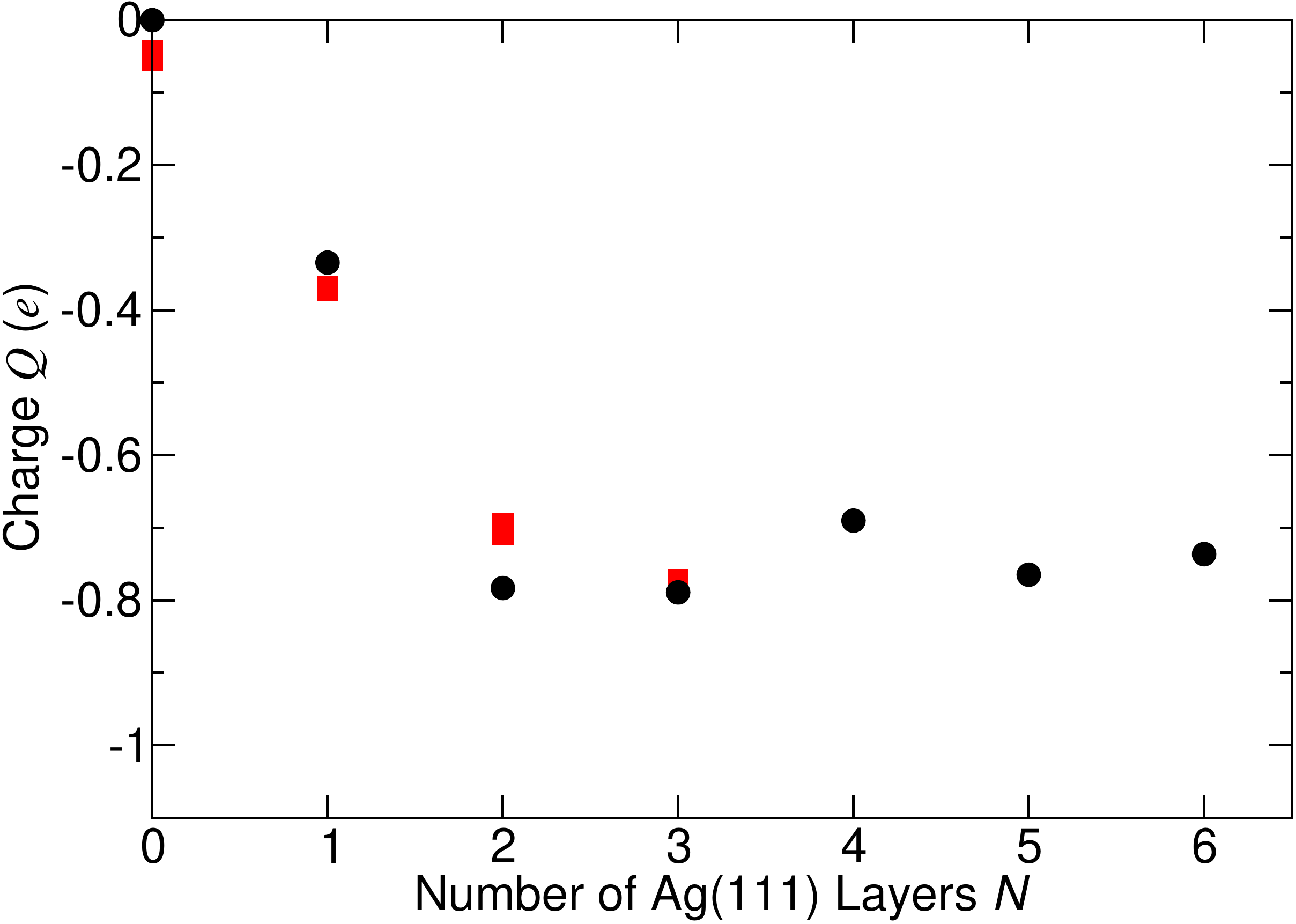}
\caption{Charge $Q$ in $e$ of PFP in a pure monolayer ({$\medbullet$}) and a 1:1 mixed PFP+CuPc monolayer ({\color{red}{$\blacksquare$}}) as a function of the number of Ag(111) atomic layers $N$, calculated via a Bader analysis.}
\label{fgr:charge_vs_layers}
{\color{JPCCBlue}{\rule{\columnwidth}{1pt}}}
\end{figure}

Figure~\ref{fgr:charge_vs_layers} shows the calculated charge $Q$ of PFP in a pure monolayer as a function of the number of layers $N$ of the Ag(111) substrate.  We find that for $N = 3$ the charge transfer to the pure PFP layer is already converged to the limit of $Q \approx -0.75$~$e$.  
This suggests a three layer Ag(111) slab should provide a good description of charge transfer from the infinite slab, at a reasonable computational cost.  Since we are mostly interested in an accurate description of charge transfer within our donor--acceptor/metal systems, we may employ a three layer Ag(111) slab model for describing the pure CuPc and the mixed 1:1 PFP+CuPc monolayers.

For the mixed 1:1 PFP+CuPc monolayer, the calculated charge of PFP increases monotonically with the number of layers $N$ of the Ag(111) substrate, as shown in Figure~\ref{fgr:charge_vs_layers}.  However, the charge transfer between CuPc and PFP remains quite small, as is seen from comparing the charge of PFP in the pure monolayer and 1:1 mixture with CuPc. 
This suggests that the effect of the Ag substrate on charge transfer between PFP and CuPc in their mixtures is quite small, and most probably within the accuracy of the calculation. For this reason, calculations of PFP and CuPc pure and mixed monolayers in vacuum may suffice to describe XPS measurements on the Ag(111) surface.


\subsection{Pure PFP and CuPc Monolayers}

As discussed in Section~\ref{sec:theoreticalmodel}, to determine a molecule's charge $Q$ from a core-level shift requires that the initial state be neutral. For this reason we have calculated the core-level shifts between a PFP monolayer in vacuum and adsorbed on a three layer Ag(111) surface with the initial state method, as depicted in Figure~\ref{fgr:initialstatemethod}.  To assess the reliability of these results we have also compared them with the XPS core-level shifts between multilayer PFP and a monolayer of PFP on Ag(111).  This is quite reasonable, since the neutral multilayer of PFP on Ag(111) should have quite similar C1s binding energies to the neutral PFP monolayer in vacuum.

\begin{figure*}
\includegraphics[width=1.5\columnwidth]{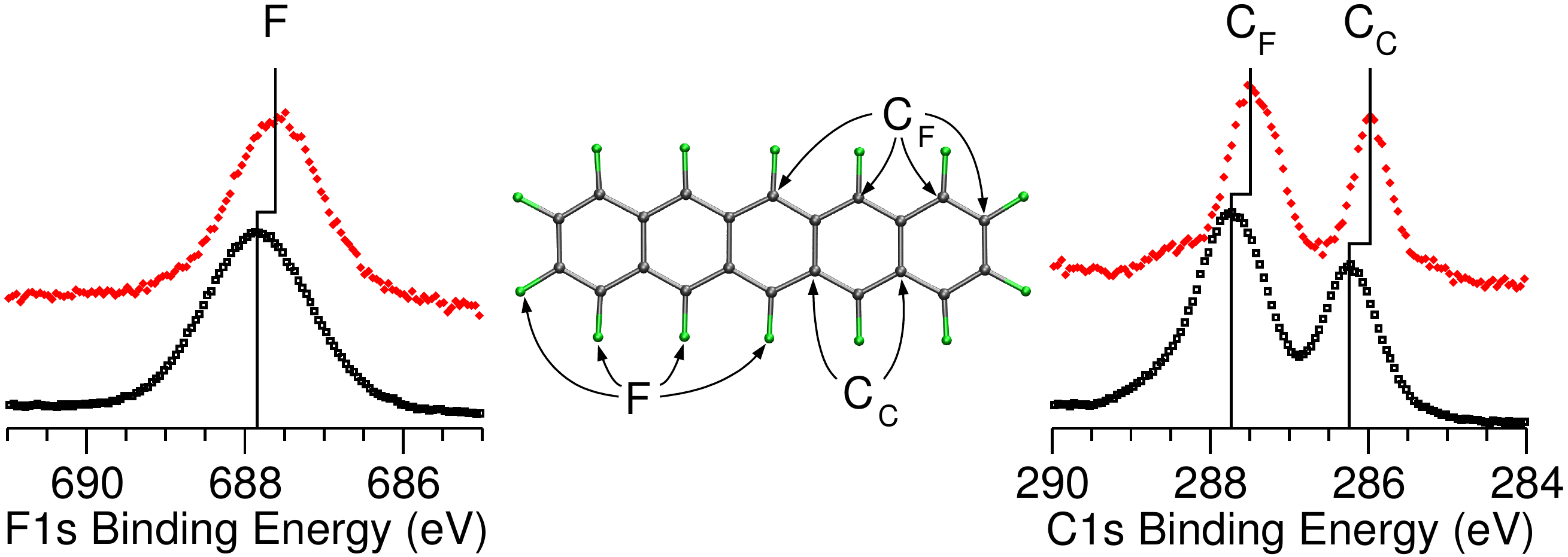}
\caption{XPS spectra for multi-layer PFP ({\footnotesize{$\square$}}) and a monolayer of PFP ({\color{red}{{\footnotesize{$\Diamondblack$}}}}) on Ag(111) of the F (left), C$_{\mathrm{F}}$ and C$_{\mathrm{C}}$ (right) atomic species, as shown schematically.  C and F atoms are depicted by gray and green balls, respectively.
}\label{fgr:MultiMono}
\end{figure*}

In Figure~\ref{fgr:MultiMono} we plot the measured XPS spectra for PFP in a multilayer ($N \gtrsim 4$) and monolayer on Ag(111).  The three different chemical environments in PFP, namely C$_{\mathrm{C}}$, C$_{\mathrm{F}}$, and F, are also depicted schematically in Figure~\ref{fgr:MultiMono}.  Specifically, for C$_{\mathrm{C}}$ we measure a core-level shift of \(\Delta E \approx -0.24\)~eV.  Small shifts to weaker binding energy such as these are often found when moving from a multilayer to a monolayer on a metal substrate, and are typically attributed to the stronger core-hole screening by the surface.  However, we will show that there is also significant charge transfer and screening of the initial state by the metal substrate in such systems.

Figure~\ref{fgr:initialstatemethod} compares the initial state binding energies, density of states, and charge transfer for a PFP monolayer in vacuum, charged, and adsorbed on a three layer Ag(111) slab.  As seen in Figure~\ref{fgr:charge_vs_layers}, the Ag slab donates a significant amount of charge  to the PFP (\(\sim -0.75\)~$e$) upon adsorption.   Charging PFP in vacuum by the same amount yields a significant C1s core-level shift to \emph{weaker} binding energies (\(\sim +1.5\)~eV) as expected.  On the other hand, at the height $h$ of PFP above the clean Ag(111) surface, the external potential shown in Figure~\ref{fgr:initialstatemethod} is strongly binding (\(\sim -1.8\)~eV), shifting the C1s core-level to \emph{stronger} binding energies.  These two competing effects cancel, yielding a small overall core-level shift of \(\Delta E \approx 0.25\)~eV.  Although this overestimates the XPS core-level shift by about 0.5~eV, this may be attributed to the substantial difference in core-hole screening between the PFP multilayer and monolayer on Ag(111), which is not accounted for in the initial state method.  On the other hand, for core-level shifts between pure and mixed monolayers, the differences in core-hole screening should be quite small, and may be neglected. 

\begin{figure*}[!t]
\centering
\includegraphics[width=0.65\textwidth]{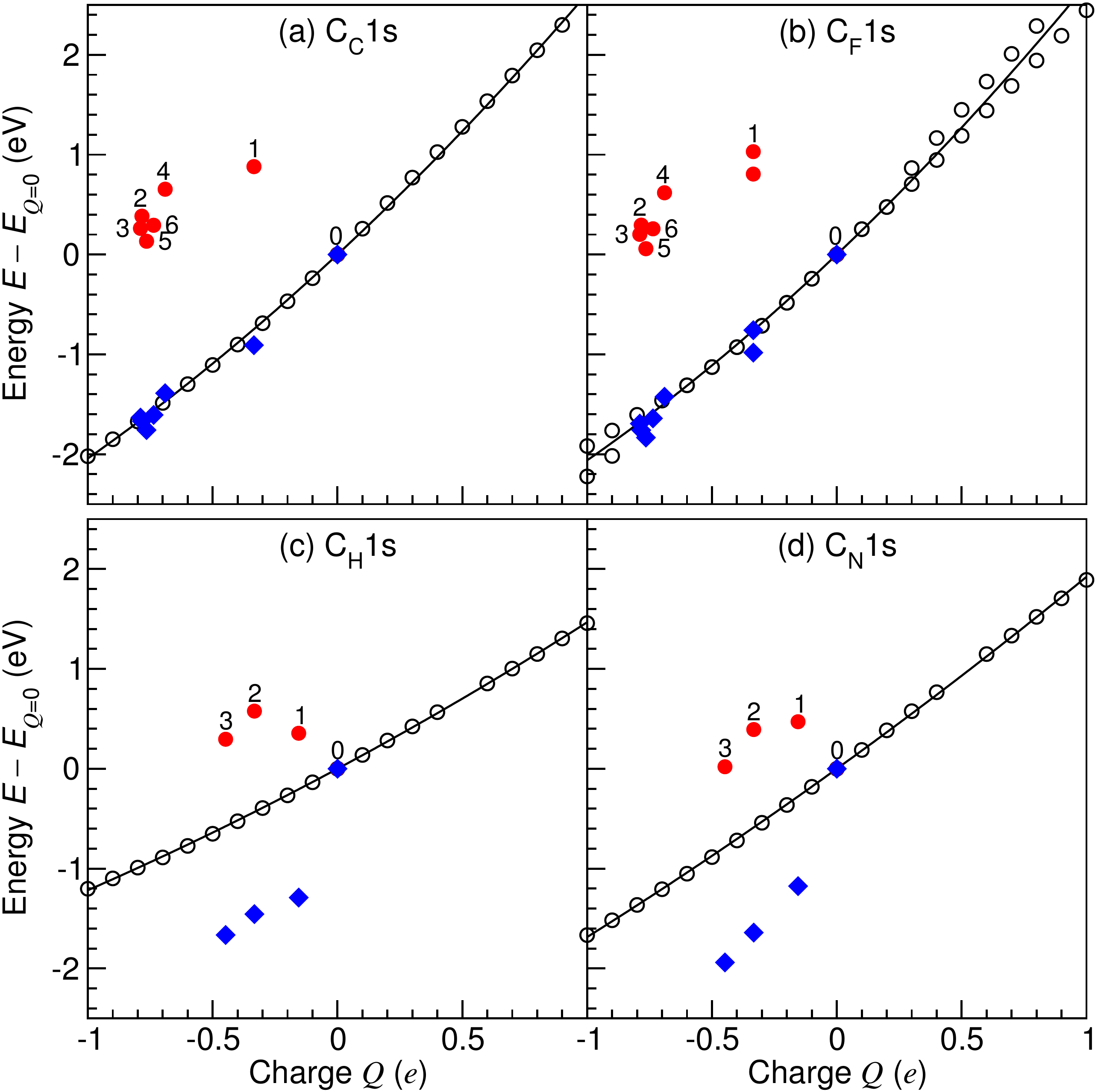}
\caption{Energy $E$ in eV versus charge $Q$ per molecule in $e$ for (a) C$_{\mathrm{C}}$ and (b) C$_{\mathrm{F}}$ atomic species in PFP and (c) C$_{\mathrm{H}}$ and (d) C$_{\mathrm{N}}$ atomic species in CuPc of the C1s level \(-\varepsilon_\mathrm{C1s}\) in vacuum ($\medcirc$), on an $N$ layer Ag(111) surface ({\color{red}{$\medbullet$}}) and after subtracting the change in external potential due to the Ag(111) surface \(-\varepsilon_\mathrm{C1s} + V\) ({\color{blue}{$\Diamondblack$}}). 
 All energies are taken relative to the binding energy of the neutral molecule $E_{Q=0}$.  A quadratic fit to the pure monolayer C1s binding energies in vacuum (\textbf{------}) is also given.
}
\label{fgr:C1sEnergiesPure}
{\color{JPCCBlue}{\rule{\textwidth}{1pt}}}
\end{figure*}

\begin{table}[!t]
\caption{\textrm{
Fitting parameters to the C$_{\textrm{C}}$, C$_{\textrm{F}}$, C$_{\textrm{H}}$, and C$_{\textrm{N}}$ 1s binding energies $\bm{-\varepsilon_\textrm{C1s}-E_{\bm{Q=0}} \approx f_{\textit{mol}}(Q) \approx \xi Q + \zeta Q^2}$ in eV for PFP and CuPc pure monolayers in vacuum, where $Q$ is the charge per molecule in \textit{e}}}\label{tbl:ModelParameters}
\begin{tabular}{ccc}
\multicolumn{3}{>{\columncolor[gray]{0.9}}c}{}\\[-3mm]
\rowcolor[gray]{0.9}level & $\xi$ (eV/$e$) & $\zeta$ (eV/$e^2$)
\\[1mm]
C$_{\textrm{C}}$1s & 2.328 & 0.285\\
C$_{\textrm{F}}$1s & 2.382 & 0.321\\
C$_{\textrm{H}}$1s & 1.341 & 0.125\\
C$_{\textrm{N}}$1s & 1.803 & 0.118\\
\end{tabular}
\end{table}

\begin{table}[!t]
\caption{\textrm{
Average effective potential \textit{V} relative to the vacuum level \textit{E}$_{\textit{vac}}$ in eV of a clean Ag(111) \textit{N} layer  surface  at the height of adsorbed PFP and CuPc pure and 1:1 mixed PFP+CuPc monolayers}}\label{tbl:VextPure}
\begin{tabular}{cccc}
\multicolumn{4}{>{\columncolor[gray]{0.9}}c}{ }\\[-3mm]
\rowcolor[gray]{0.9}Ag(111)&\multicolumn{3}{>{\columncolor[gray]{.9}}c}{$V$ (eV)}\\
\rowcolor[gray]{0.9}$N$ & \multicolumn{1}{>{\columncolor[gray]{.9}}c}{PFP} & \multicolumn{1}{>{\columncolor[gray]{.9}}c}{CuPc} & \multicolumn{1}{>{\columncolor[gray]{.9}}c}{PFP+CuPc}\\[1mm]
1 & -1.787 & -1.646 & -1.773\\
2 & -2.033 & -2.033 & -1.966\\
3 & -1.874 & -1.959 & -1.844\\
4 & -1.993 & --- & ---\\
5 & -1.890 & --- & ---\\
6 & -1.905 & --- & ---\\
\end{tabular}
{\color{JPCCBlue}{\rule{\columnwidth}{1pt}}}
\end{table}

To test the reliability of the effective potential model for core-level shifts given in \ref{eqn:DEModel}, we must first obtain a fit for $f_{\textit{mol}}(Q)$ while keeping the external environment, i.e.~the effective potential, constant.  This is accomplished by calculating core-level shifts for the monolayer in vacuum when applying an external charge $Q$ through an appropriate shift of the Fermi level.

Figure~\ref{fgr:C1sEnergiesPure} shows the calculated C1s core-level energies for a PFP monolayer in vacuum as a function of the applied charge $Q$. We find separate local maxima in the DOS \(\varepsilon_{\textrm{C1s}}\), related to the different C bonding environments or atomic species in the system, namely C$_{\textrm{C}}$ and C$_{\textrm{F}}$ as depicted schematically in Figure~\ref{fgr:MultiMono} for PFP, and C$_{\textrm{H}}$ and C$_{\textrm{N}}$ for CuPc.  For each atomic species, we find the core-level shifts are described quantitatively by \ref{eqn:fmol}, with fitting parameters $\xi$ and $\zeta$ given in Table~\ref{tbl:ModelParameters}.  Taken together, these results show that the core-level shifts are indeed linearly dependent on the charge transfer into an atom, as assumed in Section~\ref{sec:theoreticalmodel}.

\begin{figure*}[!th]
\centering
\includegraphics[width=0.65\textwidth]{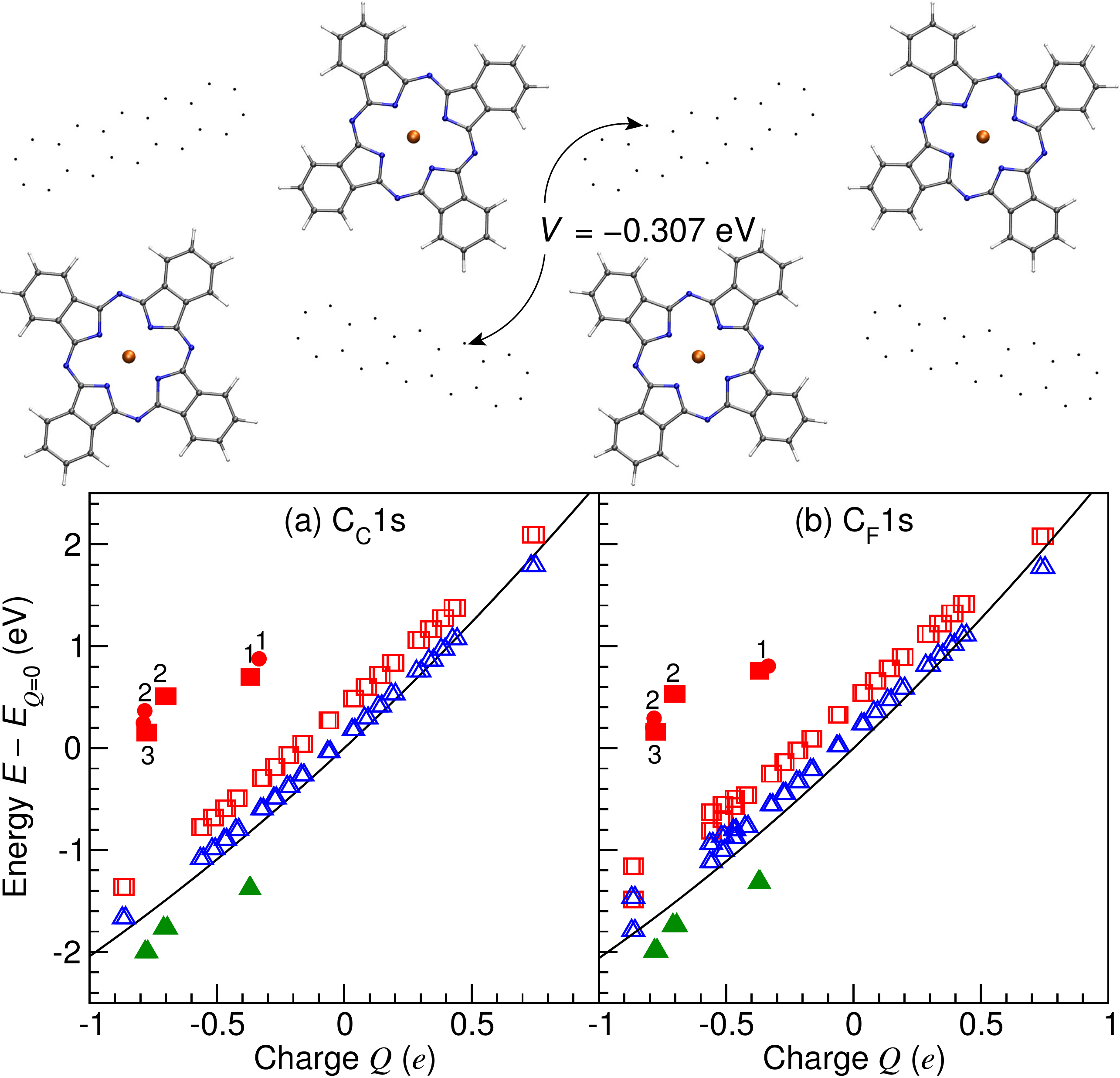}
\caption{Energy $E$ in eV versus charge $Q$ of PFP in $e$ for (a) C$_{\textrm{C}}$ and (b) C$_{\textrm{F}}$ atomic species of the C1s level \(-\varepsilon_\textrm{C1s}\) in a 1:1 mixture with CuPc in vacuum ({\color{red}{$\square$}}), on an $N$ layer Ag(111) surface ({\color{red}{$\blacksquare$}}), and after subtracting the change in external potential \(-\varepsilon_\textrm{C1s} + V\) due to the other molecules in vacuum ({\color{blue}{$\triangle$}}) and due to the $N$ layer Ag(111) surface ({\color{Green}{$\blacktriangle$}}). 
  The binding energies of the pure PFP monolayer adsorbed on an $N$ layer Ag(111) surface ({\color{red}{$\medbullet$}}) are provided for comparison.
All energies are taken relative to the binding energy of the neutral molecule $E_{Q=0}$.  A quadratic fit to the pure monolayer C1s binding energies in vacuum (\textbf{------}) is also given. The mixed 1:1 CuPc+* structure, where the average external potential ${V} \approx -0.307$~eV is calculated at the positions of the C atoms in PFP, is depicted schematically above.  C, N, H, and Cu atoms are depicted by gray, blue, white and orange balls, respectively.}
\label{fgr:C1sEnergiesMix}
{\color{JPCCBlue}{\rule{\textwidth}{1pt}}}
\end{figure*}

\begin{figure*}[!th]
\includegraphics[width=1.5\columnwidth]{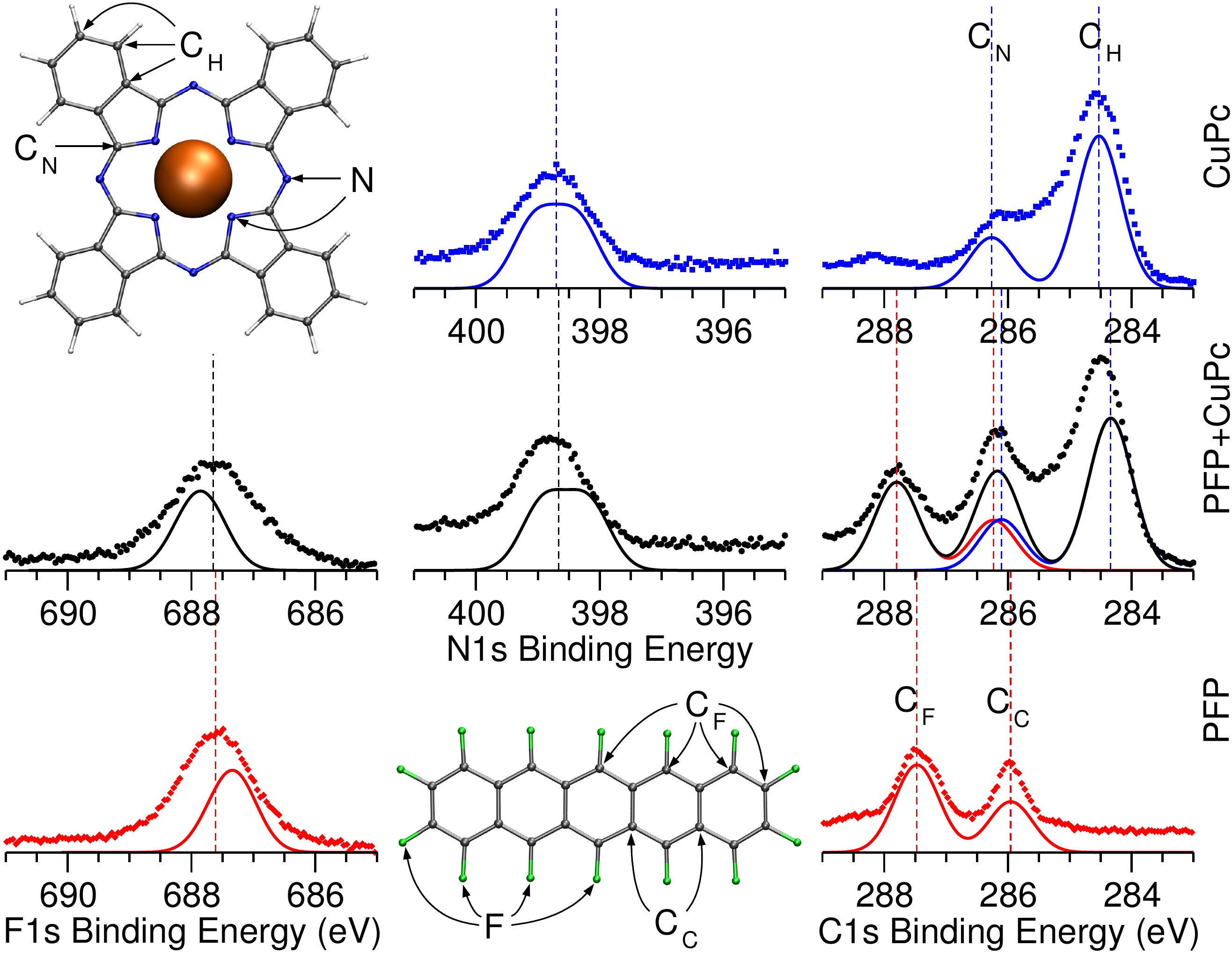}
\caption{The calculated density of states (DOS) (lines) and measured XPS spectra (symbols) for the F1s, N1s, and C1s levels versus binding energy for pure monolayers of PFP ({\color{red}{------}},{\color{red}{$\Diamondblack$}}), CuPc ({\color{blue}{------}},{\color{blue}{$\blacksquare$}}), and a 1:1 mixture of PFP+CuPc (------,$\medbullet$).  The PFP and CuPc structures, along with the four different C atomic species, C$_{\textrm{C}}$, C$_{\textrm{F}}$, C$_{\textrm{H}}$, and C$_{\textrm{N}}$, consisting of six and four symmetrically inequivalent C atoms in PFP and CuPc, respectively, are shown above.  C, F, N, H, and Cu atoms are depicted by gray, green, blue, white and orange balls, respectively.}
\label{fgr:spectraVac}
{\color{JPCCBlue}{\rule{\textwidth}{1pt}}}
\end{figure*}

Using our calculated fit to $f_{\textit{mol}}$, we may now test how well the change in screening of the core-level $g$ may be approximated by the change in the effective potential, as given in \ref{eqn:gscr} and \ref{eqn:DEModel}.  This may be accomplished by using a change in the molecule's environment, e.g.\ adsorption on an $N$ layer Ag(111) surface, to charge the molecule.

In Figure~\ref{fgr:C1sEnergiesPure} (a) and (b) we compare the core-level energies for a pure PFP monolayer adsorbed on $N$ layer Ag(111) surfaces, where \(N=1,2,3,\ldots,6\). The variation of the charge transfer from the surface to the molecule with number of layers, as shown in Figure~\ref{fgr:charge_vs_layers}, means that these calculations provide a further test of the reliability of the model for \(\Delta E\) given in \ref{eqn:DEModel}.  Little correlation is initially obvious between the C1s core-levels \(\varepsilon_{\textrm{C1s}}\) and the charge transfer \(Q\) for the pure layers on Ag(111).  However, upon removing the effect of screening, i.e.\ plotting \(-\varepsilon_{\textrm{C1s}} + V \approx f_{\textit{mol}}(Q)\), we recover the charge transfer dependence previously observed for the pure PFP layer in vacuum.

On the other hand, Figure~\ref{fgr:C1sEnergiesPure} (c) and (d) show weaker agreement when the same procedure is applied to CuPc on $N$ layer Ag(111), although the correlation with the charge transfer dependence \(f_{\textit{mol}}\) is still obtained up to a constant shift.  This suggests other contributions are present in the core-electron screening for CuPc. We attribute this to the greater screening inside the CuPc molecule and stronger interaction with the surface, due to metallic Cu---Ag chemical bonds.

Taken together, these results validate three major assumptions made in Section~\ref{sec:theoreticalmodel}.  Namely, that (1) $f_{\textit{mol}}$ is linearly dependent on the charge of an atom, (2) $f_{\textit{mol}}$ is independent of the external environment, and (3) $g$  may be reasonably approximated by the change in effective potential of the external environment for PFP, while for CuPc screening within the molecule and chemical interaction with the substrate are also important.


It should also be noted that the charge of the molecules $Q$ is directly specified for calculations of the monolayer in vacuum, while on the Ag(111) surface $Q$ is obtained from a Bader analysis.  This agreement suggests that a Bader analysis provides an excellent description of the charge transfer for these systems.

However, as discussed in Appendix~\plainref{sec:comparison}, there is a significant difference between the core-hole screening of the PFP monolayer in vacuum and on the Ag(111) surface.  This suggests that the calculated initial state core-level shifts should be shifted by \(\sim -0.4\)~eV to describe the XPS measurements.  To avoid such a discrepancy, and provide a better comparison between the calculated initial state core-level shifts and XPS measurements, we shall next compare pure and mixed monolayers of PFP and CuPc on Ag(111) in Section~\ref{sec:mixed}.

\subsection{Mixed 1:1 PFP+CuPc Monolayers}\label{sec:mixed}

As a further test of the effective potential model, we next calculate core-level shifts upon charging a 1:1 mixture of PFP and CuPc.
As shown in Figure~\ref{fgr:C1sEnergiesMix}, the mixed 1:1 PFP+CuPc monolayer in vacuum follows the $f_{\textit{mol}}(Q)$ relation up to a constant shift.  Overall, for PFP the core-level is shifted to higher binding energies ($\sim 0.3$ eV), while for CuPc it is shifted to lower binding energies ($\sim -0.2$ eV) when the two layers are mixed.  This is in near quantitative agreement with the experimental results on Ag(111), as shown in Figure~\ref{fgr:spectraVac}.


To estimate the change in external potential between the pure and mixed PFP+CuPc monolayers, we have performed separate calculations of the relaxed mixed layer geometry in vacuum with PFP removed (CuPc+*) and with CuPc removed (PFP+*).  The average effective potential at the coordinates of the C atoms in the empty sites is then calculated relative to the vacuum energy, \(V = V(\mathbf{r}_a) - E_{\textit{vac}}\), as depicted schematically in Figure~\ref{fgr:C1sEnergiesMix}.  For the CuPc+* layer we obtain a change in external potential of $\sim-0.307$~eV, which brings the core-level shifts for PFP onto the pure layer values, as seen in Figure~\ref{fgr:C1sEnergiesMix}.  This suggests that for PFP both \ref{eqn:gscr} for the screening and \ref{eqn:fmol} are valid.  Further, it shows that the charge transfer dependent portion of $\Delta E$, i.e.\ $f_{\textit{mol}}(Q)$, is independent of the external environment, and defined by the molecular environment alone.

On the other hand, for the PFP+* layer we obtain a negligible external potential shift, so that the core-level shift is overestimated by the model of \ref{eqn:DEModel}.  However, this discrepancy may again be explained by greater screening in the CuPc molecule due to the Cu metal atom.

For the mixed 1:1 PFP+CuPc monolayer on Ag(111), we have assumed the external potentials from CuPc+* and Ag(111) are additively separable, so that \(V \approx V^{\textrm{Ag(111)}} + V^{\textrm{CuPc+*}}\).  Based on the semi-quantitative agreement shown in Figure~\ref{fgr:C1sEnergiesMix} between $f_{\textit{mol}}(Q)$ and $-\varepsilon_{\textrm{C1s}} + V$, this does appear to be the case.

Figure~\ref{fgr:C1sEnergiesMix} also shows the core-level shifts between the pure PFP monolayer and the 1:1 mixture with CuPc on an $N=1,2,3$ layer Ag(111) slab. By comparing with the charge transfer for the same systems, shown in Figure~\ref{fgr:charge_vs_layers}, we find overall the core-levels shift to weaker or stronger binding energy when charge is transferred out of or into PFP, respectively.  This means the core-level shifts are strongly dependent on the number of layers in the Ag(111) surface.

Finally, in Figure~\ref{fgr:spectraVac} we directly compare the experimental XPS spectra with the total DOS for monolayers of PFP, CuPc and the mixed 1:1 PFP+CuPc monolayers in vacuum.  We find that by shifting the C$_{\textrm{F}}$1s and C$_{\textrm{H}}$1s peaks to match the pure monolayer experimental peaks for PFP and CuPc, respectively, we describe the experimental core-level shifts and relative binding energies for the pure and mixed monolayers near-quantitatively.  This suggests that the inclusion of the surface, although providing a significant charge transfer, is a nearly constant shift, so that calculations for the monolayers in vacuum remain an effective means of describing core-level shifts.

On the other hand, the requirement of separate shifts for the CuPc and PFP monolayers suggests that the details of the PFP---CuPc interactions in the mixed layer are not completely captured at the LDA level.  Further calculations including long range van der Waals type interactions may be necessary to describe both the PFP and CuPc binding energies with a single energy shift.  However, determining the charge transfer into the molecules based on the XPS core-level shifts only requires an accurate description of the effective potential, as we will show in Section~\ref{sec:chargemodel}.

\subsection{Charge Transfer Model}\label{sec:chargemodel}

\begin{figure}[!t]
\centering
\includegraphics[width=0.8\columnwidth]{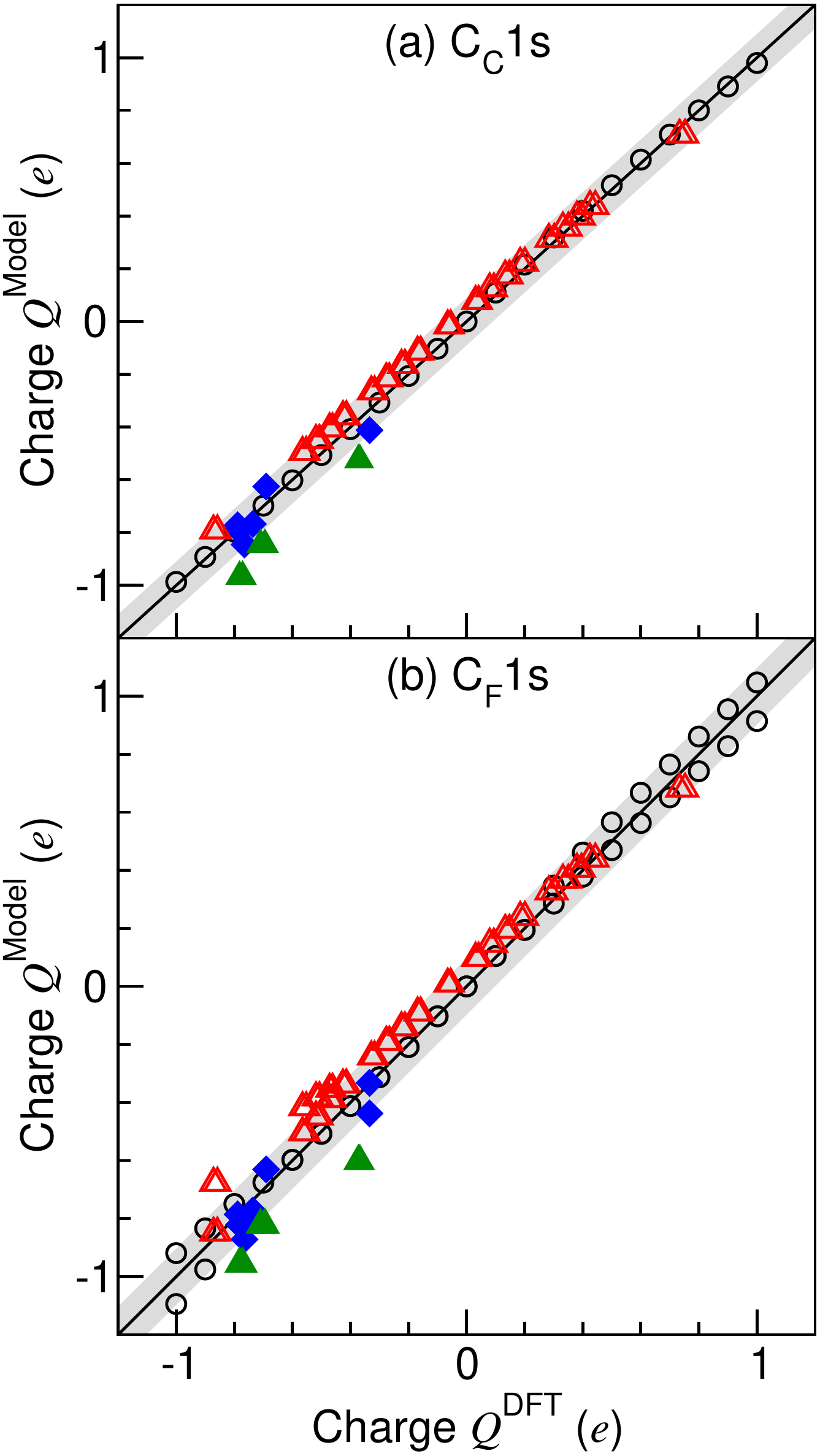}
\caption{Comparison of charge $Q$ in $e$ from an effective potential model \(Q^{\textrm{Model}}\) and from DFT calculations \(Q^\textrm{DFT}\) for PFP in vacuum ($\medcirc$), on an $N$ layer Ag(111) surface ({\color{blue}{$\Diamondblack$}}), in a 1:1 mixture with CuPc in vacuum ({\color{red}{$\triangle$}}), and on an $N$ layer Ag(111) surface ({\color{Green}{$\blacktriangle$}}).  
  The standard deviations for PFP and PFP+CuPc on Ag(111), $\sigma \approx \pm 0.09$~$e$, are shown as regions of gray.
}
\label{fgr:QModel}
{\color{JPCCBlue}{\rule{\columnwidth}{1pt}}}
\end{figure}

To determine the reliability of the effective potential model for describing charge transfer based on C1s binding energies, we next compare the calculated charge transfer $Q$ with that obtained from \ref{eqn:Qmodel}.

In Figure~\ref{fgr:QModel} we plot the calculated and model charge transfer, $Q^{\textrm{DFT}}$ and $Q^{\textrm{Model}}$ respectively, for a a pure PFP monolayer and a 1:1 mixture with CuPc monolayers in vacuum and on Ag(111).  From \ref{eqn:Qmodel}, we find for the initial state model that
\begin{equation}
Q^{\textrm{Model}} \equiv -\frac{\xi}{2\zeta} + \sqrt{\frac{\xi^2}{4\zeta^2} + \frac{-\varepsilon_{\textrm{C1s}} - E_{Q=0} + V}{\zeta}}
\end{equation}
where the model parameters \(\xi\), \(\zeta\), and \(V\) are provided in Tables~\ref{tbl:ModelParameters}, \ref{tbl:VextPure}, and Figure \ref{fgr:C1sEnergiesMix}.

We find that for PFP the charge transfer into the molecule is near-quantitatively described by the model.  Specifically, for pure PFP and its 1:1 mixture with CuPc on Ag(111), the standard deviation between $Q^{\textrm{Model}}$ and $Q^{\textrm{DFT}}$ is $\sigma \approx \pm 0.09$~$e$, as shown in Figure~\ref{fgr:QModel}.  
These results strongly support the potential use of the core-level shift relative to a molecule in vacuum to describe the charge transfer upon adsorption and molecular mixing on a metal surface.

 \begin{table}[!t]
\caption{\textrm{
Charge \textit{Q}$^{\textrm{Model}}$ in \textit{e} of PFP in a pure and mixed 1:1 PFP+CuPc monolayer on Ag(111) from an effective potential model using XSW heights \textit{h} in \AA$^{\textit{a}}$ to calculate the effective potential for Ag(111) \textit{V} relative to the vacuum level \textit{E}$_{\textit{vac}}$ in eV, combined with the XPS C1s core-level shifts $\bm{\Delta E}$ in eV$^{\textit{b}}$}}\label{tbl:Qexp}
\begin{tabular}{c@{}cr@{.}lr@{.}lr@{.}lr@{.}l}
\multicolumn{10}{>{\columncolor[gray]{0.9}}c}{ }\\[-3mm]
\rowcolor[gray]{0.9}&&\multicolumn{4}{>{\columncolor[gray]{0.9}}c}{PFP} & \multicolumn{4}{>{\columncolor[gray]{0.9}}c}{PFP+CuPc}\\
\rowcolor[gray]{0.9}&&\multicolumn{2}{>{\columncolor[gray]{.9}}c}{C$_{\textrm{C}}$1s}&\multicolumn{2}{>{\columncolor[gray]{.9}}c}{C$_{\textrm{F}}$1s}&\multicolumn{2}{>{\columncolor[gray]{.9}}c}{C$_{\textrm{C}}$1s}&\multicolumn{2}{>{\columncolor[gray]{.9}}c}{C$_{\textrm{F}}$1s}\\[1mm]
$h$ &(\AA)                    & 3&16 & 3&16 & 3&28 & 3&51\\
$V$ &(eV) & -1&62 & -1&62 & -1&51 & -1&30\\
$\Delta E$ &(eV)         & -0&26 & -0&24 & 0&00 & 0&07\\
$Q^{\textrm{Model}}$ &($e$)      & -0&91 & -0&89 & -0&87 & -0&71\\
\end{tabular}
\begin{flushleft}
$^a$XSW heights for C$_{\textrm{C}}$ and C$_{\textrm{F}}$ in PFP and PFP+CuPc on Ag(111) taken from refs.~\citenum{Duhm2010PFPAg111}~and~\citenum{XWSGoiri}, respectively. $^b$XPS core-level shifts taken relative to multilayer PFP.
\end{flushleft}
{\color{JPCCBlue}{\rule{\columnwidth}{1pt}}}
\end{table}

In Table~\ref{tbl:Qexp} we show the results of applying the effective potential model to estimate the charge transfer to PFP based on the experimental core-level shifts.  Here we have used experimental x-ray standing wave (XSW) measurements to determine the heights $h$ for C$_{\textrm{C}}$ and C$_{\textrm{F}}$ atomic species in pure PFP \cite{Duhm2010PFPAg111} and mixed PFP+CuPc \cite{XWSGoiri} monolayers on Ag(111).  Based on this data, we then use a DFT calculation for a 13 layer Ag(111) slab to determine the effective potential at a height $h$ above the surface $V(h)$.  Combining this with the XPS core-level shifts, effective potential for CuPc+* of $-0.307$~eV, and the fitting parameters for $f_{\textit{mol}}$ provided in Table~\ref{tbl:ModelParameters}, we obtain from \ref{eqn:Qmodel} the charge of PFP $Q^{\textrm{Model}}$, given in Table~\ref{tbl:Qexp}.

We find a charge of about \(-0.9\)~$e$ is donated to PFP by the Ag(111) surface, in both the pure and mixed monolayers.  This suggests there is very little net charge transfer to PFP when going from the pure monolayer to a 1:1 mixture with CuPc in the experiment.  This explains why the calculations for the monolayers in vacuum describe the XPS core-level shifts so well in Figure~\ref{fgr:spectraVac}.  It should be noted these results most probably overestimate the charge transfer when going from the multilayer to the monolayer of PFP, as the XPS core-level shifts also include differences in the strength of the core-hole screening.  As this was found to be about 0.4~eV, as discussed in Appendix \plainref{sec:comparison}, we may expect the actual charge transfer to be closer to $-0.7$~$e$, in agreement with the DFT results shown in Figure~\ref{fgr:charge_vs_layers}.  In any case, by combining the results of XPS and XSW measurements with DFT calculations, we estimate that there is a significant charge transfer to PFP upon adsorption on a Ag(111) surface, which is basically unchanged by mixing with a CuPc donor molecule.

\begin{figure}[!t]
\includegraphics[width=0.75\columnwidth]{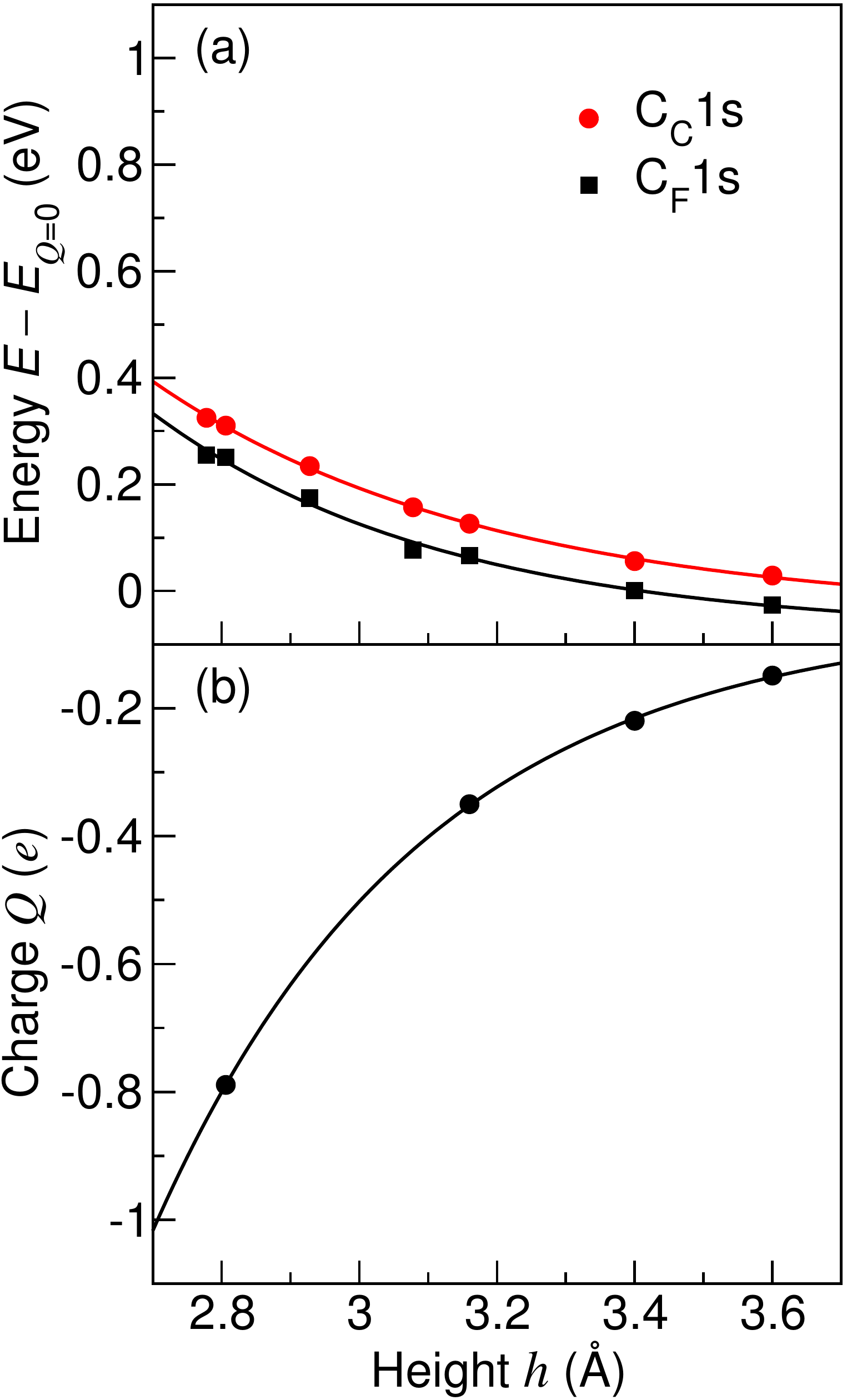}
\caption{(a) Energy $E$ of the C$_{\textrm{C}}$1s ({\color{red}{$\medbullet$}})  and C$_{\textrm{F}}$1s ({\color{black}{$\blacksquare$}}) levels in eV and (b) charge $Q$ in $e$ ($\medbullet$) of a pure PFP monolayer versus height $h$ in \AA\ above a three layer Ag(111) surface. Energies are taken relative to the binding energy of the neutral molecule $E_{Q=0}$ in vacuum. Exponential fits are provided as guides to the eye.}
\label{fgr:EC1sQvsh}
\end{figure}
 \begin{table}[!h]
\caption{\textrm{
Calculated heights \textit{h} in \AA\ of each type of atomic species in the PFP and CuPc pure and 1:1 mixed PFP+CuPc monolayers above Ag(111).}}\label{tbl:heights}
\begin{tabular}{cccc}
\multicolumn{4}{>{\columncolor[gray]{0.9}}c}{ }\\[-3mm]
\rowcolor[gray]{0.9} atomic & \multicolumn{3}{>{\columncolor[gray]{.9}}c}{$h$ (\AA)}\\
\rowcolor[gray]{0.9} species & \multicolumn{1}{>{\columncolor[gray]{.9}}c}{PFP} & \multicolumn{1}{>{\columncolor[gray]{.9}}c}{CuPc} & \multicolumn{1}{>{\columncolor[gray]{.9}}c}{PFP+CuPc}\\[1mm]
C$_{\textrm{C}}$      & 2.82 &  --- & 2.86 \\
C$_{\textrm{F}}$      & 2.80 &  --- & 2.82 \\
F                   & 2.73 &  --- & 2.72 \\
C$_{\textrm{H}}$      &  --- & 2.71 & 2.94 \\
C$_{\textrm{N}}$      &  --- & 2.80 & 2.98 \\
H                   &  --- & 2.71 & 2.90 \\
N                   &  --- & 2.82 & 3.01 \\
Cu                  &  --- & 2.73 & 2.93 \\
\end{tabular}
{\color{JPCCBlue}{\rule{\columnwidth}{1pt}}}
\end{table}

It should be noted, however, that LDA calculations typically underestimate heights of weakly adsorbed molecular monolayers on metal surfaces.  This is clearly seen by comparing the heights for PFP and CuPc pure and 1:1 mixed monolayers on Ag(111) from XSW measurements \cite{Duhm2010PFPAg111,XWSGoiri} with LDA results, as shown in Tables~\ref{tbl:Qexp} and \ref{tbl:heights}, respectively.  We find LDA calculations consistently yield heights for C in PFP at \(\sim 2.8\)~\AA\ above the Ag(111) surface.  This is in contrast to XSW measurements, which find both C$_{\textrm{C}}$ and C$_{\textrm{F}}$ atomic species at \(h\approx 3.16\)~\AA\ in the pure PFP monolayer, and much higher at $h\approx 3.28$ and 3.51~\AA, respectively, in the 1:1 mixture with CuPc, \emph{cf}.\ Table~\ref{tbl:Qexp}.  

To understand how these discrepancies may affect the reliability of the effective potential model,  we have calculated the dependence of the calculated core level shifts $\Delta E$ and charge $Q$ on the height $h$ of a pure PFP monolayer on three layer Ag(111).  This is accomplished by performing separate initial state core-level calculations and Bader analyzes after rigidly shifting the PFP monolayer to a height $h$ above the Ag(111) surface.  

Figure~\ref{fgr:EC1sQvsh} (a) shows that as the PFP monolayer is raised off the surface, the C$_{\textrm{C}}$1s and C$_{\textrm{F}}$1s binding energies decrease monotonically to the binding energy for the neutral monolayer in vacuum, $E_{Q=0}$.  Further, the dependence of the core-level shifts on the height of the molecule is rather weak, changing by less than 0.2 eV between the calculated and measured PFP heights of 2.82 and 3.16 \AA, respectively.  On the other hand, as shown in Figure~\ref{fgr:EC1sQvsh} (b), the charge $Q$ of PFP has a stronger dependence on the height $h$, with $Q \sim -0.8$ and $-0.4$~$e$ at the calculated and measured PFP heights, respectively.  As expected, we find the charge of PFP decays monotonically to zero as the monolayer is raised off the Ag(111) surface.

Taken together, these results suggest that although the charge transfer to the PFP monolayer $Q$ decreases in magnitude with increasing height $h$, this is countered by a decrease in magnitude of the external potential $V$ from the Ag(111) substrate.  In effect, changes in $f_{\textit{mol}}(Q)$ and $V$ with $h$ balance, so that the core-level shifts $\Delta E$ change rather little.  Overall, this suggests LDA initial state calculations of core-level shifts should provide a reliable description of XPS measurements, and an effective potential model based on LDA parameters may be applied to estimate charge transfer based on XPS core-level shifts and XSW heights.

\section{CONCLUSIONS}\label{sec:conclusions}

In summary, we have derived and applied an effective potential approach to describe charge transfer within a reticular donor-acceptor/metal complex based on core-level shifts.  To do so we have performed DFT calculations and XPS measurements of core-level shifts for PFP, CuPc, and mixed 1:1 PFP+CuPc layers adsorbed on Ag(111). We find that the calculated core-level shifts are described near-quantitatively in terms of the charge transfer into the molecule, and the change in external potential from the environment, which captures the effect of screening for the weakly interacting PFP molecule.

Using this model, we were able to estimate the charge transfer into a molecule using the experimental core-level shift relative to the pure multilayer crystal, and the calculated change in effective potential due to the other molecules and the metallic substrate.  This provides a novel method for the direct assessment of charge transfer in weakly interacting molecule--substrate systems via XPS measurements and routine DFT calculations.  However, further study is needed for other donor--acceptor/metal systems, e.g.\ PEN or FCuPc on Cu or Au, to fully assess the applicability of the effective potential approach.

\appendix
\titleformat{\section}{\bfseries\sffamily\color{JPCCBlue}\normalsize}{\Large$\blacksquare$\normalsize~APPENDIX~\thesection.~}{0pt}{}

\section{STM}\label{sec:stm}

The STM image in Figure~\ref{fgr:STM} (b) has been obtained from the DFT electronic structure calculations using the Tersoff-Hamann approximation~\cite{stm}, as implemented in \textsc{gpaw}~\cite{gpaw1,gpaw2}. In this approach the current $I$ at a position $\mathbf{r}$ is given by
\begin{equation}
\displaystyle
I(\mathbf{r})\approx CV\sum_{n\mathbf{k}} \exp\left(-\frac{(\varepsilon_{n\mathbf{k}}-\varepsilon_F)^2}{\Delta^2}\right)\frac{\psi_{n\mathbf{k}}(\mathbf{r})\psi_{n\mathbf{k}}^*(\mathbf{r})}{N_{\mathbf{k}}}
\label{eqn:stme}
\end{equation}
where $\psi_{n\mathbf{k}}$ is the $n^{\textrm{th}}$ Kohn-Sham wave function at $\mathbf{k}$-point $\mathbf{k}$ with eigenenergy $\varepsilon_{n\mathbf{k}}$, $C$ is a prefactor that depends on the density of states (DOS), surface work
function and radius of the tip, $V$ is the potential of the sample with respect to the tip,  $\Delta \approx 0.1$~eV is the electronic width in the calculation, $N_{\mathbf{k}}$ is the weight of $\mathbf{k}$-point $\mathbf{k}$, and $\varepsilon_F$ is the Fermi energy.  As a Fermi energy we have used the energy corresponding to the highest occupied  Kohn-Sham state at the $\Gamma$ point of the clean Ag(111) surfaces.

\section{FINAL STATE METHODOLOGIES}\label{sec:finalstatemethods}
\begin{figure}[!t]
\centering
\includegraphics[width=0.85\columnwidth]{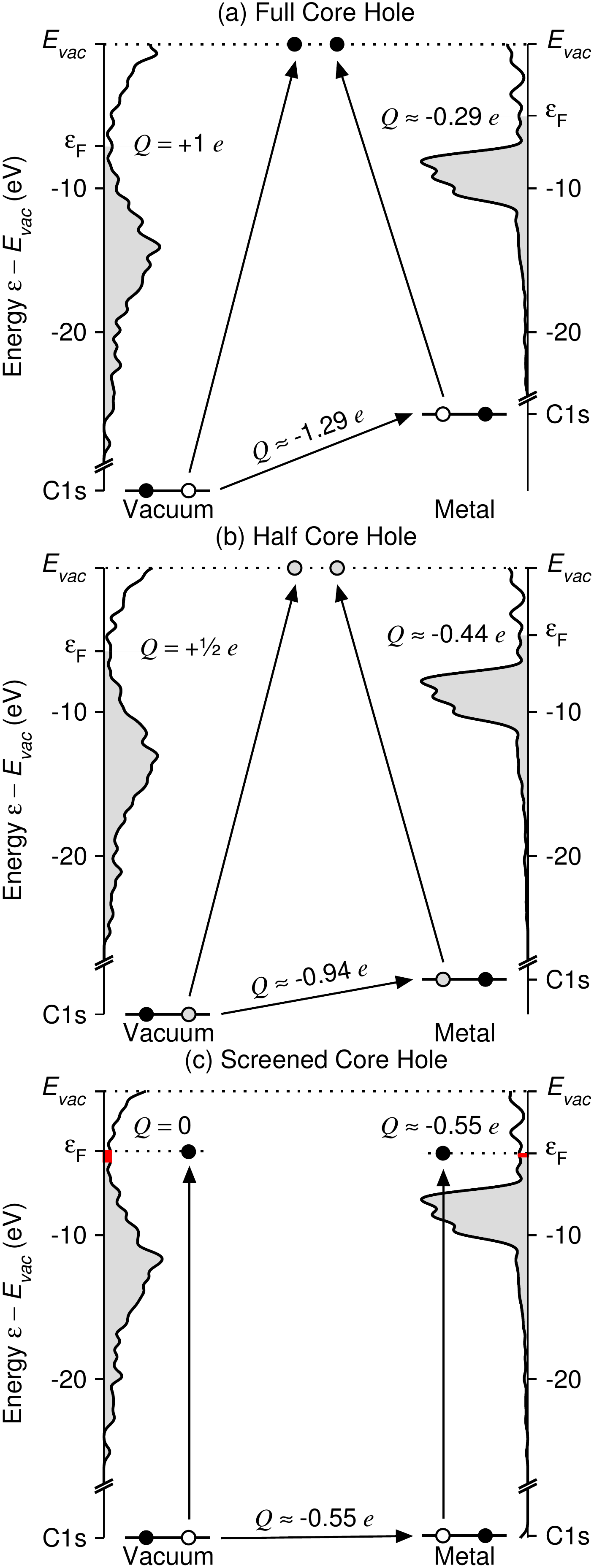}
\caption{Schematics depicting the (a) full core-hole, (b) half core-hole, and (c) screened core-hole final state methods for calculating core-hole energy shifts, as applied to a PFP monolayer in vacuum, and adsorbed on a Ag(111) three layer surface. For each method the calculated density of states (DOS) for the final state, charge of the molecule $Q$ in the final state, and C1s energy shifts are shown.  Occupation of the DOS is denoted by grey regions, with charge added to screen the core-hole marked in red. Note  the DOS for the PFP monolayers in vacuum were increased by a factor of five for clarity.
}
\label{fgr:finalstatemethods}
{\color{JPCCBlue}{\rule{\columnwidth}{1pt}}}
\end{figure}

To determine the reliability of the initial state approach for core-level shifts due to molecular adsorption on a metallic substrate, we also calculated the binding energies using three different final state methods, as depicted schematically in Figure~\ref{fgr:finalstatemethods}.  In each final state method the binding energy of the C1s level $E$ is obtained from the total energy difference between the final state with the C1s level unoccupied, and the initial ground state.  In other words,
\begin{equation}
E \approx E_{f} - E_{GS}
\end{equation}
where \(E_{f}\) is the total energy of the final state with the photoelectron ejected from the C1s level, and \(E_{GS}\) is the total energy of the system in the ground state.  In this way the ability of the environment to screen a core-hole is included explicitly through the final state.

The simplest method to model the final state is using a full core-hole, as depicted schematically in Figure~\ref{fgr:finalstatemethods}(a).  In this case a special pseudopotential is employed for one of the C atoms, which has one electron removed from the C1s level, i.e.\ a full core-hole.  Within this model, the ejected photoelectron is then assumed to be ejected into the vacuum.  This requires a separate DFT calculation for each inequivalent C chemical environment in the system, since the total energy with a full core-hole \(E_{1ch}\) is required.

A major advantage of this method is that, by taking a difference of total energies, the absolute binding energies obtained are often within 1--2\% of the experimental values.  However, this still means discrepancies of more than 2 eV are often found with experiment, with the value obtained highly dependent on the xc-functional employed.  For this reason, the results still need to be ``shifted'' when compared with experiment.

This method also has the drawback that the binding energy is very strongly dependent on the system's ability to screen the core-hole, which will depend on the number of weakly bound valence electrons in the system.  For a minimal unit cell containing only one molecule, the screening due to polarization of neighboring neutral molecules is completely neglected.  Since the number of electrons available to screen is an order of magnitude higher for a molecule on a metal substrate compared to a molecule in vacuum, this may lead to a significant overestimation of the 
core-level shifts.

To address this issue, the half core-hole method, shown in Figure~\ref{fgr:finalstatemethods}(b), has also been tested.  This again requires a separate total energy calculation \(E_{\nicefrac{1}{2}ch}\), for each inequivalent chemical environment, with a special pseudopotential with half an electron removed from the C1s level applied to one of the C atoms.  The photoelectron is again assumed to be moved to the vacuum level, but the core-hole is now assumed to be partially screened.

This type of calculation has the advantage that it requires less screening of the core-hole by the environment.  However, the binding energies from half core-hole calculations are typically about half those of full core-hole calculations, so that as with the initial state method, only energy shifts should be compared with experiment.

The third final state method considered herein involves a screened core-hole, as seen in Figure~\ref{fgr:finalstatemethods}(c).  This method differs from the full core-hole method only in that the photoelectron is moved to the bottom of the conduction band, via a charged calculation.  By charging the system, the core-hole may be fully screened, whether the molecule is isolated or adsorbed on a metal surface.

We find this method yields core-level shifts which agree qualitatively with the initial state method, suggesting that effects due to screening of the final state should be similar in vacuum or on a metal substrate.  However, as the photoelectron is not ejected from the system, such a calculation more properly describes an x-ray adsorption spectroscopy (XAS) experiment.  In this way, such a calculation will only describe XPS results in the limit where XAS and XPS experiments are in quantitative agreement.

As depicted in Figure~\ref{fgr:finalstatemethods}, the final state methods for calculating the binding energy $E$ of the C1s core level involve the following procedure:
\begin{enumerate}
\item Perform a DFT structural relaxation of the molecular system to obtain the ground state energy $E_{GS}$.
\item Calculate the final state energy $E_f$ using a core-hole PAW pseudopotential for one atom in the molecule with either:
\begin{enumerate}
\item a full core-hole pseudopotential $Q = +1$~$e$  (full core-hole method),
\item a half core-hole pseudopotential $Q = +\nicefrac{1}{2}$~$e$ (half core-hole method), or
\item a full core-hole pseudopotential with an electron added to the conduction band $Q = 0$ (screened core-hole method).
\end{enumerate}
\item Calculate the final state binding energies using $E = E_f - E_{GS}$.
\end{enumerate}

\section{PFP FINAL STATE RESULTS}\label{sec:comparison}

When a final state full core-hole calculation is performed, as shown in Figure~\ref{fgr:finalstatemethods}(a), we find a quite significant charge transfer from the Ag(111) surface to PFP ($\sim -1.29$~$e$), which is almost double that of the initial state.  This transfer reflects the ability of the Ag(111) surface to completely screen the core-hole through charge donation.

On the other hand, the PFP layer in vacuum is unable to move significant amounts of charge, since only a single PFP molecule is included within the unit cell.  As a result, the core-hole shift is significantly overestimated by the full core-hole method ($\sim -5.3$~eV compared to $-0.24$~eV from XPS).  This difference between the molecule on the surface and in the vacuum may be partially addressed through the inclusion of uncharged molecules in the vacuum unit cell.  However, this quickly becomes computationally unfeasible, and limits the comparability of the relevant calculations.

For a half core-hole calculation, as seen in Figure~\ref{fgr:finalstatemethods}(b), we obtain a charge transfer from the Ag surface to PFP which is between the initial state and full core-hole results ($\sim -0.94$~$e$).  The calculated core-level shift is also between the full core-hole and initial state results ($\sim -2.5$~eV), again overestimating the XPS results. This suggests further screening of the core-hole is necessary to describe core-level shifts upon adsorption on a metallic substrate.

From Figure~\ref{fgr:finalstatemethods}(c) we see that when the core-hole is completely screened, both the charge transfer to PFP ($\sim -0.55$~$e$) and core-hole shift ($\sim -0.14$~eV) agree semi-quantitatively with the initial state method (\emph{cf}.\ Figure~\ref{fgr:initialstatemethod}).  More importantly, the core-hole shift is within 0.1 eV of the XPS measurements.  We may thus conclude that a full screening of the core-hole is required to describe core-hole shifts between molecules in vacuum and adsorbed on a metal surface.  However, as mentioned previously, a screened full core-hole calculation more correctly describes an XAS experiment.

For this reason, the discrepancies we observe between the screened core-hole and initial state results may be attributed to differences in the final state core-hole screening of the PFP monolayer in vacuum and on the Ag(111) surface.  From this we may estimate that the core-hole screening by the Ag substrate is \(\sim -0.4\)~eV.  However, when both environments are monolayers on a surface, e.g.\ core-level shifts between pure and mixed monolayers on Ag(111) at similar heights above the surface with molecules of comparable polarizabilities, differences in core-hole screening should be negligible.

Overall, these comparisons suggest that from the point of view of both computational feasibility and robustness/accuracy of results, the initial state method may be preferred for core shifts upon adsorption on a metal substrate.

\titleformat{\section}{\bfseries\sffamily\color{JPCCBlue}\normalsize}{\thesection.~}{0pt}{\Large$\blacksquare$\normalsize~}
\section*{AUTHOR INFORMATION}
\subsubsection*{Corresponding Authors}
\noindent *E-mail: duncan.mowbray@gmail.com (D.J.M.).\\ *E-mail: angel.rubio@ehu.es (A.R.).
\subsubsection*{Notes} 
\noindent The authors declare no competing financial interest.
\section*{ACKNOWLEDGEMENTS}
We acknowledge funding by the European Research Council Advanced Grant DYNamo (ERC-2010-AdG-267374),
Spanish MICINN (FIS2010-21282-C02-01, MAT2010-21156-C03-01, -C03-03, and PIB2010US-00652), ``Grupos Consolidados UPV/EHU del Gobierno Vasco'' (IT-319-07, IT-257-07), ACI-Promociona (ACI2009-1036) and European Community e-I3 ETSF project (Contract No. 211956). JLC acknowledges funding through the Mexican CONACyT program. DJM acknowledges funding through the Spanish ``Juan de la Cierva'' program (JCI-2010-08156).

\bibliography{bibliography}

\end{document}